\documentclass[10pt,conference]{IEEEtran}
\usepackage{cite}
\usepackage{amsmath,amssymb,amsfonts}
\usepackage{algorithmic}
\usepackage{graphicx}
\usepackage{textcomp}
\usepackage{xcolor}
\usepackage{xargs}
\usepackage[colorinlistoftodos,prependcaption,textsize=small]{todonotes}
\usepackage[hyphens]{url}

\def\BibTeX{{\rm B\kern-.05em{\sc i\kern-.025em b}\kern-.08em
    T\kern-.1667em\lower.7ex\hbox{E}\kern-.125emX}}

\usepackage{setspace}
\usepackage{algorithm}
\usepackage{subfig} 
\usepackage{booktabs}
\usepackage{pifont}

\usepackage{todonotes}

\let\oldding\ding
\renewcommand{\ding}[2][1]{\scalebox{#1}{\oldding{#2}}}

\definecolor{ao(english)}{rgb}{0.0, 0.5, 0.0}
\definecolor{ao}{rgb}{0.007, 0.520, 0.867}
\definecolor{gfored}{rgb}{0.580, 0.050, 0.211}
\definecolor{carminepink}{rgb}{0.92, 0.3, 0.26}
\definecolor{OrangeRed}{rgb}{0.93, 0.8, 0.75}

\definecolor{amber}{rgb}{1.0, 0.75, 0.0}

\newcommand{\del}[1]{\textcolor{red}{}} 

\definecolor{darkcyan}{rgb}{0.0, 0.55, 0.55}

\newif\ifsubmission
\submissionfalse
\ifsubmission
\newcommand{\ieyp}[2][]{}
\newcommand{\nisc}[1]{}

\else
\newcommand{\ieyp}[2][]{\todo[backgroundcolor=carminepink!20,bordercolor=purple,inline,#1]{#2}}
\newcommand{\nisc}[1]{\textcolor{darkcyan}{[\textbf{@nisa:}#1]}}

\fi

\newcommandx{\atanote}[2][1=]{\todo[linecolor=blue,backgroundcolor=blue!25,bordercolor=blue,#1,size=\footnotesize]{#2}}
\newcommandx{\inlineatanote}[2][1=]{\todo[inline,linecolor=blue,backgroundcolor=blue!25,bordercolor=blue,#1,size=\small]{#2}}

\title{TuRaN: True Random Number Generation Using Supply Voltage Underscaling in  SRAMs} 
\newcommand{\affilETH}[0]{\textsuperscript{\S}}
\newcommand{\affilETU}[0]{\textsuperscript{$\dagger$}}
\newcommand{\affilSRG}[0]{\textsuperscript{$*$}}
\author{
{İsmail Emir Yüksel\affilETH\affilETU}\qquad%
{Ataberk Olgun\affilETH}\qquad%
{Behzad Salami\affilSRG}\qquad
{F. Nisa Bostanc{\i}\affilETH}\qquad
{Yahya Can Tu\u{g}rul\affilETH\affilETU}\qquad\\
{A. Giray Ya\u{g}l{\i}kç{\i}\affilETH}\qquad%
{Nika Mansouri Ghiasi\affilETH}\qquad
{Onur Mutlu\affilETH}\qquad
{O\u{g}uz Ergin\affilETU}\qquad\vspace{-3mm}\\\\
{\vspace{-3mm}\affilETH \emph{ETH Z{\"u}rich}} \qquad \affilETU \emph{TOBB University of Economics and Technology} \qquad \affilSRG  \emph{SAFARI Research Group}%
}

\makeatletter
\g@addto@macro{\normalsize}{%
  \setlength{\abovedisplayskip}{3pt plus 0.5pt minus 1pt}
  \setlength{\belowdisplayskip}{3pt plus 0.5pt minus 1pt}
  \setlength{\abovedisplayshortskip}{0pt}
  \setlength{\belowdisplayshortskip}{0pt}
  \setlength{\intextsep}{4pt plus 1pt minus 1pt}
  \setlength{\textfloatsep}{4pt plus 1pt minus 1pt}
  \setlength{\skip\footins}{5pt plus 1pt minus 1pt}}
\makeatother

\usepackage{titlesec}
\titlespacing\section{0pt}{2pt plus 1pt minus 1pt}{3pt plus 1pt minus 2pt}
\titlespacing\subsection{0pt}{2pt plus 1pt minus 1pt}{3pt plus 1pt minus 2pt}
\titlespacing\subsubsection{0pt}{2pt plus 1pt minus 1pt}{3pt plus 1pt minus 2pt}

\begin{document}
\maketitle
\thispagestyle{plain}
\pagestyle{plain}


\begin{abstract}
True random number generators (TRNGs) rely on unpredictable physical entropy sources such as electrical noise, thermal noise, and clock jitters. However, not all computing devices are equipped with dedicated hardware to extract entropy from these sources.~Thus, it is costly to provide true random number generation capability to computing systems via dedicated hardware. 

Prior works propose SRAM-based TRNGs that extract entropy from SRAM arrays. SRAM arrays are widely used in a majority of specialized or general-purpose chips that perform computation to store data inside the chip. Thus, SRAM-based TRNGs present a low-cost alternative to dedicated hardware TRNGs. However, existing SRAM-based TRNGs suffer from 1) low TRNG throughput, 2) high energy consumption, 3) high TRNG latency, and 4) the inability to generate true random numbers continuously, which limits the application space of SRAM-based TRNGs.

Our goal in this paper is to design an SRAM-based TRNG that overcomes these four key limitations and thus, extends the application space of SRAM-based TRNGs. To this end, we propose TuRaN, a new \emph{high-throughput}, \emph{energy-efficient}, and \emph{low-latency} SRAM-based TRNG that can sustain \emph{continuous operation}. TuRaN leverages the key observation that accessing SRAM cells results in random access failures when the supply voltage is reduced below the manufacturer-recommended supply voltage. TuRaN generates random numbers at high throughput by repeatedly accessing SRAM cells  with reduced supply voltage and post-processing the resulting random faults using the SHA-256 hash function.

To demonstrate the feasibility of TuRaN, we conduct SPICE simulations on different process nodes and analyze the potential of access failure for use as an entropy source. We verify and support our simulation results by conducting real-world experiments on two commercial off-the-shelf FPGA boards. We evaluate the quality of the random numbers generated by TuRaN using the widely-adopted NIST standard randomness tests and observe that TuRaN passes all tests. TuRaN generates true random numbers with (i) an average (maximum) throughput of $1.6Gbps$ ($1.812Gbps$), (ii) $0.11nJ/bit$ energy consumption, and (iii) $278.46\mu s$ latency. TuRaN outperforms the state-of-the-art SRAM-based TRNGs by $2.26\times$, $5.09\times$, and $5.39\times$ in terms of throughput, energy efficiency, and latency, respectively.
\end{abstract}
\section{Introduction}
True random number generators (TRNGs) sample random physical phenomena (e.g., electrical noise~\cite{gong2019true,huang2014real}, atmospheric noise~\cite{zbilut2000recurrence, rahman2014ti}, thermal noise~\cite{srinivasan20102}, clock jitter~\cite{cherkaoui2013very}, noise in a compact memory~\cite{puglisi2018random}) to generate non-deterministic truly-random numbers. Random number sequences generated by TRNGs are unpredictable and irreproducible, since the source of entropy is non-deterministic. Therefore, security-critical applications use TRNGs to guarantee secure operation, as the random values generated by TRNGs do not depend on a seed value that can compromise the system security when predicted.

Modern systems use dedicated hardware TRNGs to provide true random numbers to applications.  
However, not all modern computing devices have dedicated hardware TRNGs (e.g., IoTs and mobile systems~\cite{holcomb2008power, holcomb2007initial}). To enable true random numbers in these devices, prior works propose TRNG mechanisms that use existing memory devices such as SRAMs \cite{vanderLeest2012,zhang2020improved,wang2020aging,li2015pufkey,holcomb2008power,holcomb2007initial}, DRAMs~\cite{kim2019d, olgun2021quac,talukder2019exploiting}, FLASH memories\cite{chakraborty2020true,wang2012flash}. Among these devices, SRAM (i) exists in most commodity systems even where other devices do not (e.g., RFID tag circuits~\cite{holcomb2007initial}) and (ii) is more secure as it is on-chip and does not require any off-chip link to transfer the generated random numbers to the computing unit. These advantages make SRAM a promising TRNG substrate.

Prior works on SRAM-based TRNGs~\cite{vanderLeest2012,zhang2020improved,kiamehr2017leveraging,rahman2016enhancing,sadhu2020sstrng,wang2020aging,yeh2019self,holcomb2007initial,clark2018sram,wang2020long,li2015pufkey,holcomb2008power} use start-up values as a source of entropy to generate random numbers. The start-up values of some SRAM cells settle to an unpredictable value depending on the environmental noise (e.g. temperature and voltage fluctuations) at each power-up. Unfortunately, existing SRAM-based TRNGs that rely on start-up values suffer from four key drawbacks: they~\textit{(i)} cannot sustain continuous operation and generate true random numbers with ~\textit{(ii)} low-throughput, ~\textit{(iii)} high-latency, and ~\textit{(iv)} high energy consumption, compared to the other memory-based TRNGs~\cite{kim2019d,olgun2021quac,talukder2019exploiting,ferdaus2021true}.  

\textbf{Our goal} in this paper is to overcome these drawbacks and develop 
 an SRAM-based TRNG that can be practically implemented in commodity devices while continuously providing high-throughput true random numbers with low-latency and low energy consumption.

To meet our goal, we propose a new technique to generate true random numbers in SRAM devices by underscaling the supply voltage of an SRAM device. Underscaling the supply voltage of the SRAM blocks below the manufacturer-recommended margin and accessing the SRAM cells using the nominal latency violates the required access latency, and thus causes an access failure~\cite{chen2005modeling}. We observe that not all access failures are deterministic and reading certain reduced-voltage SRAM cells induce metastability in SRAM sense amplifiers which causes sense amplifiers to sample random data.


To this end, we propose TuRaN, a new SRAM-based TRNG that leverages the access failures in SRAM cells as the source of entropy to generate true random numbers by aggressively underscaling SRAM supply voltage and post-processing the resulting errors using a cryptographic hash function. TuRaN consists of three steps: 1) experimentally identifying the SRAM rows that have high entropy under voltage underscaling operation using a low-cost profiling step as a one-time process, 2) performing read operation on the previously-identified high entropy rows and, 3) post-processing the result of the read operation using SHA-256 cryptographic hash function and generates true random numbers.

We verify the randomness of failure mechanism~\textit{i.e., access failure} that TuRaN leverages using detailed circuit-level simulations. We show that regardless of process node 1) SRAM devices are inherently susceptible to access failures, and 2) access failures can be used as a source of entropy as it exhibits randomness. To support our simulation-based observations, we conduct FPGA-based experiments.

We perform our real-world experiments and characterization on two identical samples of the Xilinx ZC702 FPGA board~\cite{zc702board} with 560 blocks of SRAMs in total. We analyze each SRAM row's entropy under four operating parameters: (i) voltage, (ii) data pattern, (iii) frequency, and (iv) temperature. We observe that the randomness caused by access failures in reduced-voltage SRAMs does not only occur in simulation environment but also occurs in commodity SRAM chips. Therefore, we expect that TuRaN is a reliable TRNG mechanism that is applicable for a wide range of SRAM devices.

We evaluate TuRaN using 560 real SRAM chips in four aspects: quality (i.e., randomness), throughput, energy, and latency. We use the NIST STS to validate the randomness of TuRaN's output and observe that random numbers generated by TuRaN pass all NIST STS tests. Our empirical results show that TuRaN generates true random numbers with the average (maximum) throughput of $1.6 Gbps$ ($1.812 Gbps$). TuRaN consumes $0.11nJ$ energy per true random bit while having $278.46\mu s$ 256-bit true random number generation latency. 

We integrate TuRaN into L1 data cache and L2 cache in a modern computing system~\cite{wikichipcascade}. We use Drowsy Cache~\cite{flautner2002drowsy} to enable TuRaN, as Drowsy Cache allows us to scale the voltage level of each cache line at negligible cost. To generate true random numbers, TuRaN takes advantage of the idleness in data cache hierarchy. To leverage the idle time intervals in caches, TuRaN follows a two step approach. First, when a cache have enough idle cycles to generate random numbers, TuRaN underscales the voltage of a previously-identified cache line that has the highest entropy and reads the sense amplifier result to obtain a random bitstream. Second, TuRaN performs the SHA-256 function using the CPU to avoid any additional area overhead. We evaluate TuRaN by running applications from SPEC2006~\cite{henning2006spec} benchmark suite on a realistic system modelled using the gem5~\cite{binkert2011gem5} system simulator and observe that TuRaN generates true random numbers in the L1 data cache (L2 cache) with an average throughput of  $4.03Gbps$ ($10.95Gbps$) and has $0.00165mm^2$ ($0.0111mm^2$) chip area overhead which is 0.0066\% (0.0444\%) of a core die area of modern high-end
CPU~\cite{wikichipcascade}) while having $4.86\%$ ($1.92\%$) performance degradation on average.

Our contributions are as follows:
\begin{itemize}
\item We introduce TuRaN, a new SRAM-based TRNG that leverages SRAMs for extracting true random numbers by performing aggressive voltage underscaling below the safe margin. To our knowledge, this study is the first work to use the voltage underscaling technique for generating true random numbers in SRAMs.
\item To evaluate the potential of true random number generation using SRAMs, we characterize SRAM access failures under four different operating parameters: voltage, data pattern, frequency and temperature using 560 SRAM blocks embedded in FPGAs.  
\item We experimentally evaluate that TuRaN is a high-quality TRNG using the standard NIST STS for randomness and show that random bitstreams extracted with TuRaN pass all tests.
\item We show that TuRaN \textit{(i)} maintains its continuous operation, \textit{(ii)} achieves $2.26\times$ higher throughput, \textit{(iii)} consumes $5.09\times$ less energy, and \textit{(iv)} has $5.39\times$ lower latency, compared to the state-of-the-art SRAM-based TRNGs,
\item We study system integration of TuRaN and show that TuRaN can generate true random numbers in the L1 data cache (the L2 cache) at $4.03Gbps$ ($10.95Gbps$) throughput while incurring a negligible area overhead of $0.00165mm^2$ ($0.0111mm^2$) and $4.86\%$ ($1.92\%$) performance degradation, on average.
\end{itemize}

\section{Background}
\subsection{Static Random Access Memory (SRAM)}
SRAMs are widely used in many computing systems as a register file, cache, branch predictor, and on-chip buffer memory. 
There are many types of SRAM bit cells with different bit topologies. In this section, we focus on six transistor SRAM (6T-SRAM) cells, as it is the conventional design and typically used in commodity devices due to its superior robustness and packing density characteristics~\cite{pavlov2008cmos}.
\subsubsection{SRAM Block Organization}
\begin{figure}[ht]
  \centering
  \includegraphics[width=0.8\linewidth]{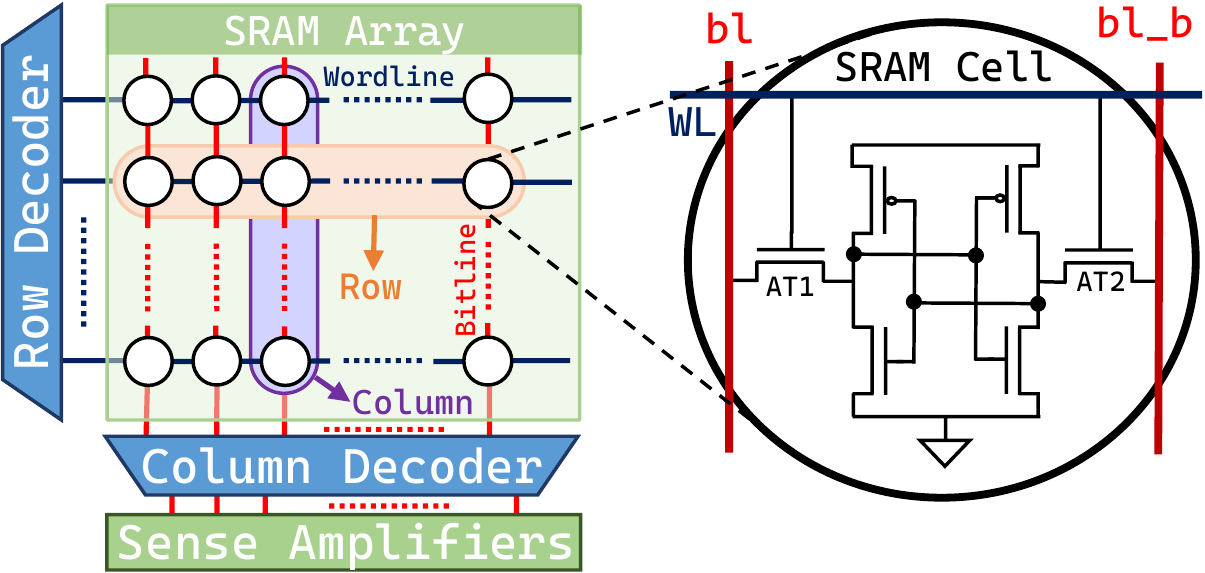}
  \caption{SRAM block and cell organization}
  \label{fig:sram_block}
\end{figure}

An SRAM block consists of an array of SRAM cells along with row and column circuitry. Figure~\ref{fig:sram_block} shows an example of an SRAM block structure and a 6T SRAM cell. 

An SRAM row is a set of SRAM cells that share a common wire, called a~\emph{word line}. A column of SRAM cells is connected to the same wire, called a~\emph{bit line}. 
A row decoder decodes the address of the accessed row and enables the corresponding word line. The column circuitry consists of the column decoder, precharge circuit, and sense amplifier. Column decoders allow sharing of a single sense amplifier among columns to perform read or write operations for only a subset of the cells in a row. A precharge circuit is used to set the voltage level of the two bit lines of the columns to the supply voltage (VDD). A sense amplifier amplifies the small voltage difference in the bitlines and produces a digital output, logic-0 or logic-1.
A conventional 6T SRAM cell consists of six transistors. The cell is composed of two identical CMOS inverters connected in a loop using four transistors. Remaining two transistors called access transistors (AT1 and AT2). Access transistors connect the bit lines (\emph{bl} and \emph{bl\_b}) and word line (WL) to the cell for read/write operations.
\subsubsection{SRAM Read Operation}
\label{sec:sram_op}
SRAM read operations consist of two steps: first in the cell and then in the sense amplifier.

\begin{figure}[ht]
  \centering
  \includegraphics[width=\linewidth]{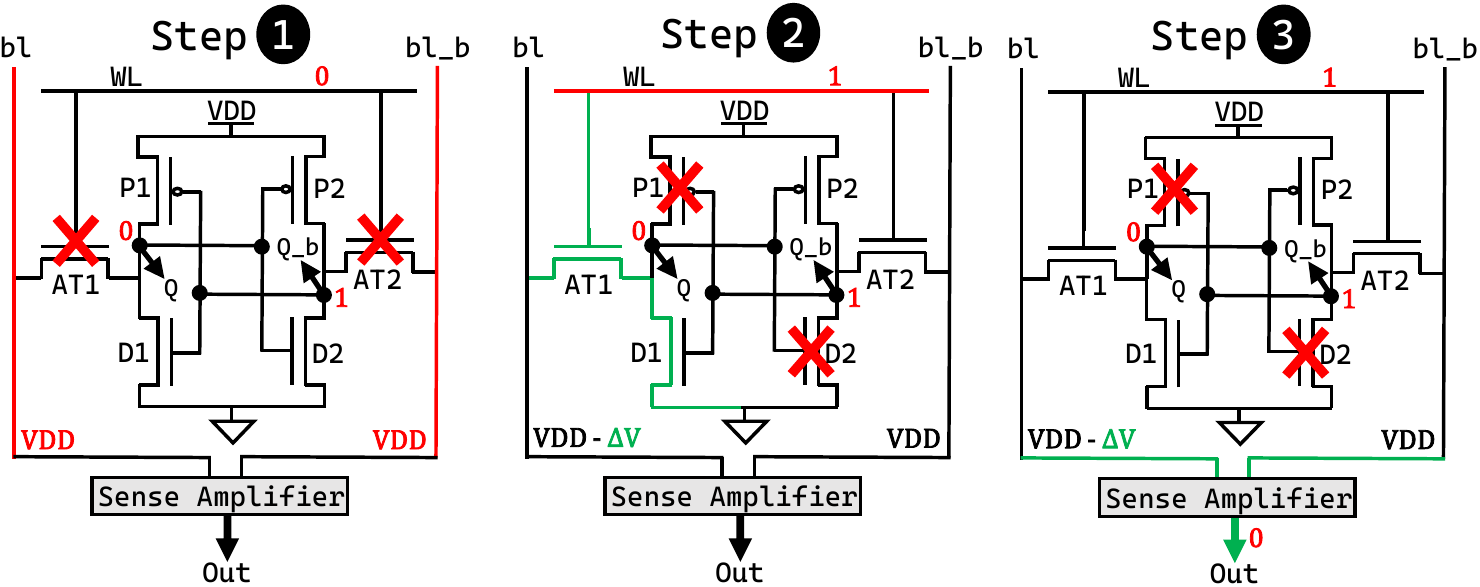}
  \caption{Read operation in an SRAM cell}
  \label{fig:read_op}
\end{figure}

\textbf{Cell Operations.} Figure~\ref{fig:read_op} illustrates the read operation in the cell of this example in three steps.  Without losing generality, assume the SRAM cell (node Q) stores \mbox{logic-0}. Hence, Q\_b is \mbox{logic-1}. Prior to initiating a read operation, the bit lines are precharged to VDD. AT1 and AT2 are both closed as their gates have 0V. (Step~\ding[1.4]{182}). 

Next in the Step~\ding[1.4]{183}, the word line is raised, and the read operation of a cell starts. In this step, AT1 and AT2 connect the cell to the precharged bit lines. P1 and D2 are both closed as their gates have 0V. Since Q stores logic-0, D1 creates a path (green line) to the ground which results in the voltage of \emph{bl} to shift towards 0V (i.e., VDD-$\Delta$V). 

In the Step~\ding[1.4]{184}, the sense amplifier detects the small voltage difference of two bit lines (\emph{bl, bl\_b}) without waiting for \emph{bl} to be fully discharged and captures the output value as logic-0.

\textbf{Sense Amplifier Operations.} Sense amplifiers in SRAM blocks sense the small analog differential voltage in the bit lines. Thus, this mechanism reduces the latency and energy consumption by saving the delay of waiting for a full bit line swing. A conventional latch-type sense amplifier initially sets its inputs to the precharge voltage level. When an SRAM cell is read and its bit lines are discharged, creating a sufficient differential voltage. The sense amplifier latches the differential voltage on bit lines by triggering the sense amplifier enable signal~\cite{pavlov2005design,pavlov2008cmos}. Shortly after that, accessed columns connect to the sense amplifier by issuing the column multiplexer signal to prevent the bit lines from being discharged by the sense amplifier~\cite{weste1985principles, ishibashi2011low}.
\subsection{Supply Voltage Underscaling}
Supply voltage underscaling below the safe margin (\textit{i.e.,} undervolting), is an effective technique to primarily save power because the total power consumption of any underlying hardware is directly related to its supply voltage \cite{updahyay2014low,azam2010variability}. By applying this technique, dynamic power consumption can be reduced quadratically. 
Voltage underscaling is studied in a wide variety of computing systems and memory devices, such as CPUs~\cite{papadimitriou2019adaptive, papadimitriou2017harnessing, gottel2021scrooge}, GPUs~\cite{leng2013gpuwattch, leng2015safe, zou2018voltage, leng2015gpu}, FPGAs~\cite{salami2018comprehensive,salami2020experimental}, DRAMs~\cite{chang2017understanding, koppula2019eden, david2011memory}, SRAMs~\cite{yang2017approximate, yang2017sram,yuksel2021mors}, HBMs~\cite{larimi2021understanding}, and NAND Flash memories~\cite{cai2018reliability, micheloni2013inside, cai2015read, cai2012error, cai2017error, cai2013threshold, cai2013error, cai2017vulnerabilities}. In addition to power savings, voltage underscaling is used for software-based fault attacks, combined with frequency scaling \cite{yang2005power,chen2021voltpillager,murdock2020plundervolt,tang2017clkscrew,kenjar2020v0ltpwn}. In commercial devices, voltage underscaling can be performed safely to some extent without affecting the accuracy of systems. However, aggressive voltage underscaling without changing the operating frequency may cause timing faults due to the increasing circuit delay, called critical region~\cite{salami2018comprehensive}. Further voltage underscaling below the minimum operating voltage systems stop operating~\cite{salami2018comprehensive,salami2020experimental,yuksel2021mors,larimi2021understanding}.

\subsubsection{Voltage Underscaling-based Faults in SRAM}
\label{sec:undervolt_faults}
A mismatch in the strength between the SRAM cell's transistors caused by the random process variations can lead to a failure during the cell operations~\cite{chen2005modeling,agarwal2006statistical,khan2010trends}. SRAM failures can be classified into four main categories: (i) read failure, (ii) write failure, (iii) access failure, and (iv) hold failure~\cite{chen2005modeling, kim2020sram}. Read failures destroy (\emph{i.e.,} flip) the data in the cell while read operation is performed. Write failures occur when the write operation fails to write the desired value into the cell. Access failures happen during the sensing operation and do not affect the SRAM cell's data because of an increase in the cell access time. Hold failures occur when the cell is not accessed for a time interval and the value is destroyed due to charge leakage.

Aggressive voltage underscaling may induce these failures during the read operation in SRAMs. At nominal voltage SRAM can quickly generate reliable voltage difference between the bit lines to ensure that the SRAM works properly~\cite{kim2020sram}. However, as we reduce the supply voltage, two things can happen which lead to failure during read operation: 1) an access failure, the voltage drop rate of the bit lines decreases which leads sense amplifier to sample incorrectly and causes access failure and 2) a read failure, the cell's value can be flipped due to the insufficient charge of the cell and the capacitance difference between cells and bit lines.

\subsection{True Random Number Generators}
True random number generators (TRNGs) rely on specialized hardware rather than computing algorithms.~Typical unpredictable sources of TRNGs are based on non-deterministic physical processes, such as thermal noise~\cite{laurenciu2015low}, jitter in clocks \cite{fischer2002true}, random telegraph noise (RTN)~\cite{brown2018low} and metastable oscillation of latches~\cite{vasyltsov2008fast}.
TRNGs sample these random physical phenomena to generate statistically uncorrelated and independent bits (i.e., bitstreams). Random number sequences generated using TRNGs do not depend on a \emph{seed} value as opposed to pseudo-random number generators (PRNGs). TRNGs typically sample biased entropy sources. The output bitstreams of a TRNG often contain a higher proportion of either logic-1 or logic-0 values. Post-processing methods are used to remove bias in TRNG bitstreams, at the cost of reduced throughput and increased latency. Post-processing methods range from simple functions (e.g., the von Neumann Corrector \cite{jun1999intel}) to cryptographic hash functions (e.g., SHA-256) with varying rates of post-processing capabilities.  

\section{Motivation and Goal}\label{sec:motivation}
True random number generators (TRNGs) are indispensable parts of various modern security-critical systems, especially cryptographic applications such as session and temporary key generation to initialize secure and private communications, secured servers, VPN access, and authentication-based applications~\cite{aysu2015end,chatterjee2017puf,sadeghi2010enhancing,che2008random}. These applications base their security on the stability and unpredictability of random numbers. A failure in the RNG part of the devices due to an adversary attack can jeopardize the security of the whole system. Prior works~\cite{chamon2021deterministic,vsimka2006active} show that systems that have poor-quality (i.e., predictable) RNGs can be significantly affected by RNG attacks. Therefore, high-quality RNGs are essential as a countermeasure against hardware attacks to maintain the security of systems.

TRNGs with high-throughput and low-latency are becoming a necessity for modern commodity devices, in particular, secure data-centric systems~\cite{qin2016things,schreckling2013kynoid,shao2008pdcs,lin2019high}. These systems are often equipped with dedicated TRNG hardware to be able to sustain their secure operations without degrading their performance. Many prior works propose different hardware-based TRNGs for such systems, including ring oscillator-based~\cite{vasyltsov2008fast,liu2016low}, chaos-based~\cite{avarouglu2015novel,galajda2006chaos,drutarovsky2007robust}, and delay chain-based~\cite{danger2009high,grujic2018closer} TRNGs. However, these hardware-based TRNGs have the following constraints: they (i) are not feasible for commodity systems because they need additional and high-complexity hardware, or (ii) cannot provide random numbers with high-throughput at low-latency. To address these issues several memory-based (e.g., DRAMs, SRAMs, NVMs) TRNGs are proposed~\cite{rashid2020true,ferdaus2021true,cambou2021trngs,kim2019d,holcomb2007initial,chakraborty2020true} as they are prevalently in use throughout a wide range of computing systems. 

SRAM has two major advantages over other memory devices: (i) it is more secure because it does not require an off-chip transfer to send the generated random bits to the CPU (i.e., SRAM is an on-chip component), and (ii) SRAM is used in every CMOS-based systems and can provide true random numbers in many devices (e.g., RFID tag circuits~\cite{holcomb2007initial}, large-scale systems~\cite{bauke2007random}). Hence, SRAM-based TRNGs offer a substrate to enable true random number applications for commodity systems.

Prior SRAM-based TRNGs~\cite{vanderLeest2012,zhang2020improved,kiamehr2017leveraging,rahman2016enhancing,sadhu2020sstrng,wang2020aging,yeh2019self,holcomb2007initial,clark2018sram,wang2020long,li2015pufkey} only use the start-up \textit{i.e., initial} values in SRAM cells that are observed immediately after the SRAM device is powered on. At the power-up state, the initial value of an SRAM cell may differ due to the process variation. Prior work~\cite{cortez2012modeling} shows that 5\%-15\% of all SRAM cells are partially-skewed and less than 5\% of them exhibit high randomness. These cells that behave randomly are used as an entropy source of true random number generation. 

These works propose viable TRNG mechanisms, however, because they depend on expensive power-up cycles and have low entropy in SRAM cells, they suffer from four major weaknesses that make them impractical for real system integration:
existing SRAM-based TRNGs (i) cannot generate true random numbers in a streaming manner, (ii) incur high latency due to the period of power-up cycle (e.g. ${\sim}250\:ms$ \cite{zhang2020improved}), (iii) cannot generate true random numbers with high-throughput~\cite{cortez2012modeling} and (iv) consume high energy for low-power energy-efficient devices.

We posit based on our analysis of prior works that an SRAM-based TRNG needs to satisfy the following properties:
\begin{itemize}
\item It must consistently generate true random numbers in a streaming manner with high-throughput at low-latency for high-performance systems.
\item It needs to consume low energy to generate true random numbers for energy-efficient devices.
\item It must be practical to implement on commodity devices from low-power edge devices to the high-throughput large-scale systems. 
\end{itemize}
Our goal in this work is to design an SRAM-based TRNG that meets all the above specifications to generate truly random numbers that are widely available for commodity devices.
\section{Faults in Reduced-Voltage SRAMs}

Supply voltage underscaling can be a promising technique for using SRAM devices as an entropy source. By leveraging this technique for a TRNG mechanism based on SRAMs, we can 1) generate true random numbers continuously since it does not require the power up cycle, 2) reduce the energy consumption of true random number generation because it naturally leverages the voltage underscaling and 3) can be easily implemented on commodity devices as recent modern systems are already equipped with a dedicated voltage controller for SRAM-based memories (e.g. caches in CPUs~\cite{hammarlund2014haswell}, on-chip memories in FPGA~\cite{xilinxdcac}).

As we discuss in Section~\ref{sec:undervolt_faults}, there are two failure mechanisms in SRAM devices that can occur during read operation: (i) read failures and (ii) access failures. To study the potential of leveraging the supply voltage underscaling technique for an SRAM-based TRNG, we analyze the read and access failure mechanisms at reduced-voltage levels using SPICE simulations.
We model 6T SRAM circuitry in 16, 22, 32, and 45 nm process nodes using PTM transistor models~\cite{cao2009ptm}. Our SRAM models consist of a 6T SRAM cell, a precharge circuitry, a write driver, and a sense amplifier. To study process variation and model SRAM cells with different transistor characteristic, we perform a Monte Carlo simulation with 1000 iterations by applying a Gaussian distribution with a 20\% standard deviation, which randomly shifts the transistor threshold voltages and bitline capacitances. We conduct this experiment in five steps: 1) write logic-1 to SRAM cell in the nominal voltage, 2) underscale the supply voltage, 3) perform read operation to SRAM cell, 4) increase the voltage back to nominal level, and 5) perform read operation at the nominal voltage. To distinguish which mechanism causes the bit failure (\emph{i.e.,} read failure or access failure), we track the cells' data during both underscaled voltage and nominal voltage by read operations. If a read failure is occurred, we expect an error at third and fifth steps as the read failures destruct the data stored in the cell. If we only observe failure in third step, this indicates that the bit flip does not occur inside the SRAM cell (\emph{i.e.,} the failure mechanism does not destroy the cell's data) but induce the error only during the sensing operation at underscaled voltage level, an access failure.

\begin{figure}[ht]
\centering
\captionsetup[subfigure]{justification=centering}
\subfloat[Read Failure]{{\includegraphics[width=0.5\linewidth]{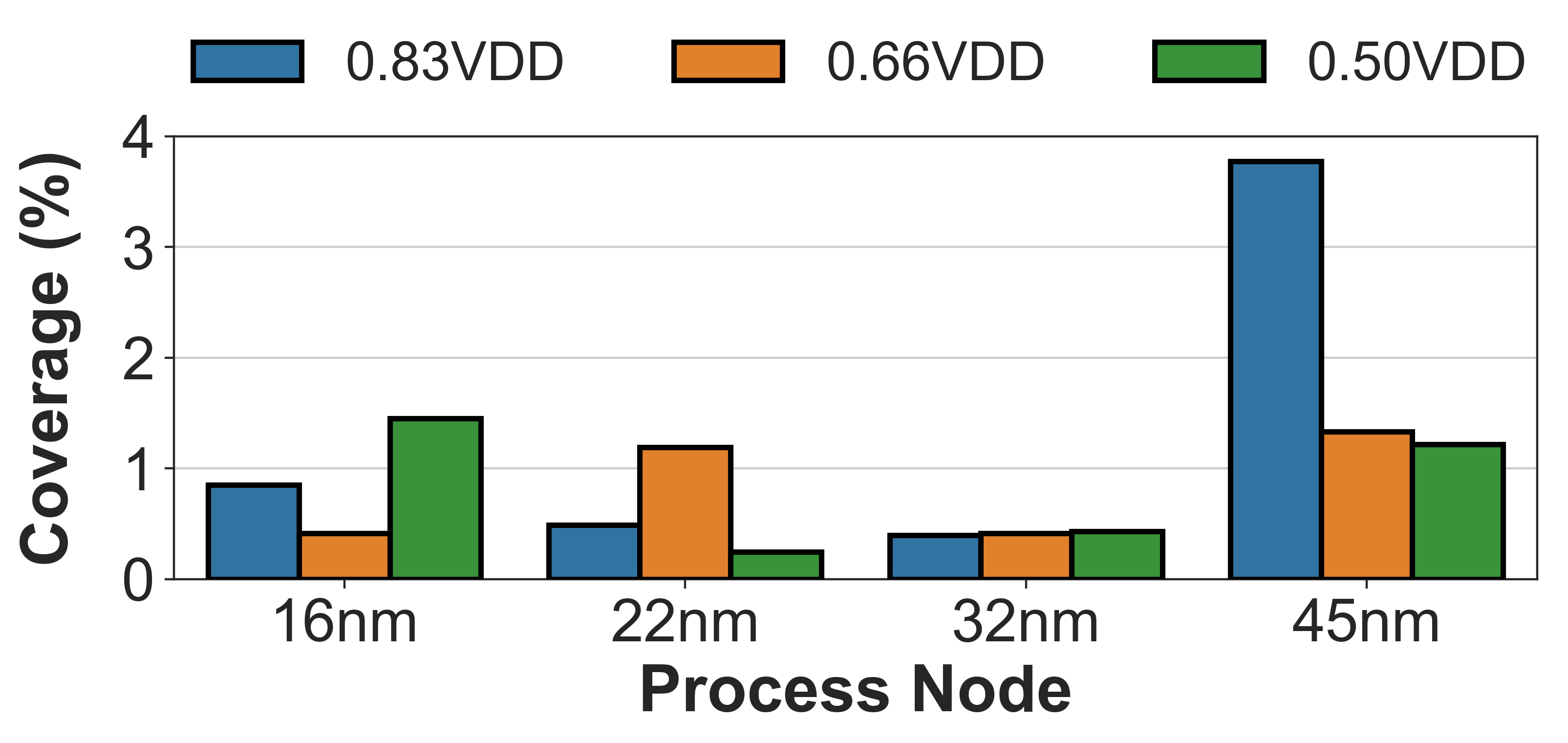} }\label{subfig:pv_read}}
\subfloat[Access Failure]{{\includegraphics[width=0.5\linewidth]{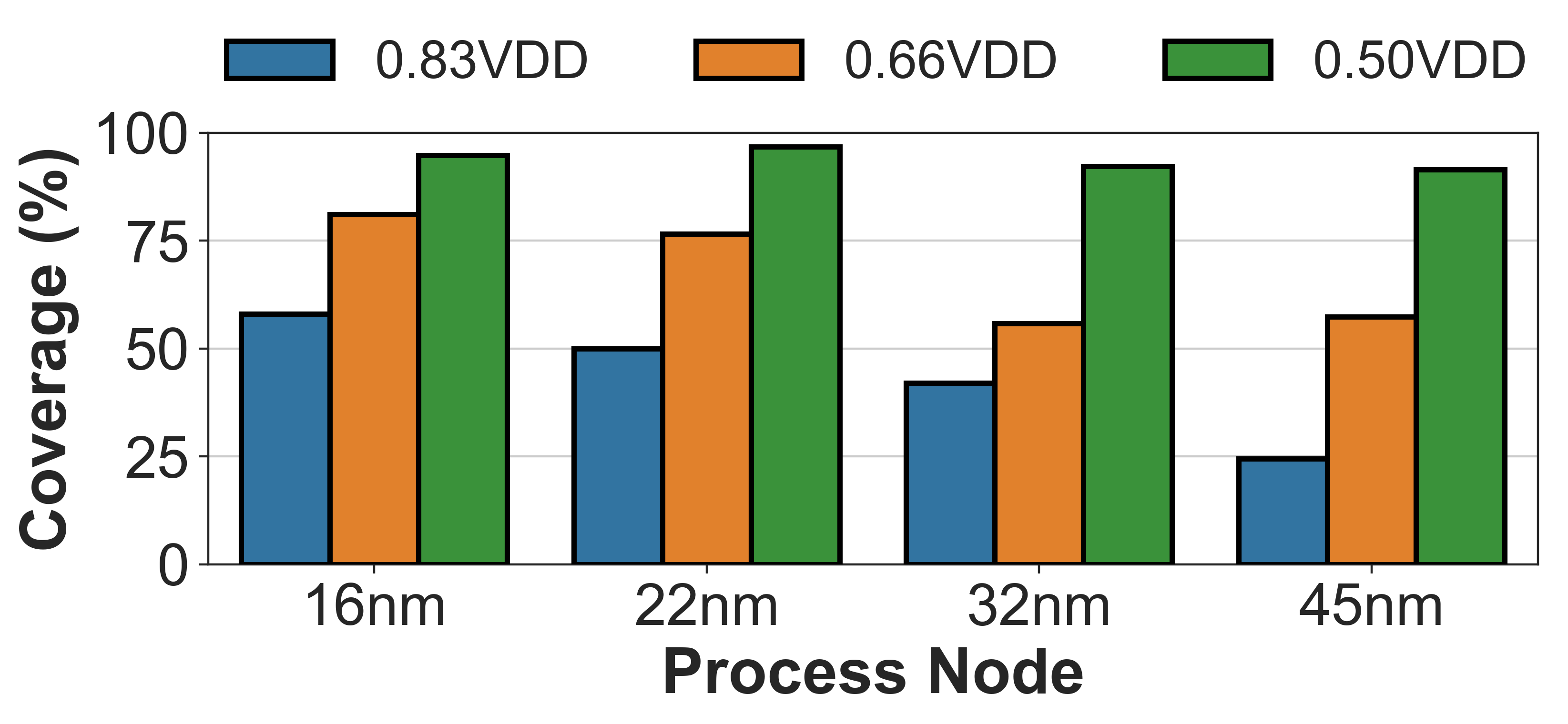} }\label{subfig:pv_access}}
\caption{Maximum and average entropy of 32-bit block for different voltage levels in each SRAM chip.}
\label{fig:pv_failures}
\end{figure}

Figure~\ref{fig:pv_failures} depicts the coverage of cells with read failures and access failures for different voltage levels across process nodes where the nominal voltage of an SRAM is VDD. We make three observation from Figure~\ref{fig:pv_failures}. First, read errors occur in less than 4\% of cells and are not significantly affected by voltage underscaling and process node. Second, access failures occur more frequently than read failures for all tested process nodes and reduced-voltage levels. Third, access failure rate increases as we decrease the supply voltage level. We conclude that access failures occur more than read failures and are affected by voltage underscaling. Thus, access failures can be a good randomness source when the voltage underscaling is leveraged.

To study the potential of access failures as a source of entropy, we conduct another SPICE simulation. We select a 16nm SRAM model as it is the most recent technology among other tested process nodes. We apply a randomized noise to our transistor model using the state-of-the-art methodology also used in prior works~\cite{camsari2017implementing, katz2008robust, rangarajan2017spin}. We run our model in a Monte Carlo simulation at 0.5VDD supply voltage for 1000 times. From this experiment, we observe access failures at 69.17\% of runs. This indicates that access failures are not deterministic such that they do not occur all the time. We hypothesize that this is because access failures cause sense amplifier to amplify a differential voltage below the reliable sensing margin, as the prior work reports~\cite{bhargava2015robust}. Hence, the sense amplifier indeterminately samples the differential voltage (probability of 50\% to VDD or GND).

Since the access failures of exact SRAM cells causes the sense amplifiers to sample the bitline voltage randomly, we expect that many different types of SRAM devices exhibit randomness. This observation indicates that the access failure can be a good candidate as a source of entropy. Hence, we study real-world experiments to analyze how the access failure in real SRAM chips behaves and whether it is on par with the simulation results.

\section{Characterization of Randomness \\in Reduced-Voltage SRAMs}
\label{sec:characterization}
We experimentally study the randomness characteristics of SRAM cells across different operating voltage and frequency levels, and data patterns. We conduct experiments on SRAM-based on-chip memories in FPGA boards. This platform enables us \textit{(i)} to manipulate voltage rails (for voltage underscaling), including individually adjusting the supply voltage of SRAM blocks, \textit{(ii}) to have the flexibility to operate in different frequency levels, and \textit{(iii)} to experiment with a large number of SRAM blocks.
\subsection{Characterization Methodology}
\label{sec:characterization-methodology}
We perform our experiments on two identical samples of Xilinx Zynq ZC702 FPGA boards (XC7Z020-CLG484-1)~\cite{zc702board} fabricated at a 28nm technology node. These boards enable independent voltage scaling of SRAM blocks via separate voltage rails.
Each FPGA board has 560 SRAM blocks and each SRAM block consists of 1024 rows and 16 columns total of 16Kbits. The nominal supply voltage of SRAM blocks, set by the manufacturer~\cite{xilinxdcac}, is $1V$.

To perform voltage underscaling, we use Power Management Bus (PMBus) standard~\cite{pmbus} to manipulate voltage rails. These rails are fully configurable and addressable by using PMBus. We use a processing system (PS) to configure PMBus through the $I^2C$ interface. We use the same interface to monitor the operating temperature, current of the corresponding voltage rail, and power consumption. In this study, we focus on $V_{CCBRAM}$, the supply voltage of SRAM-based on-chip memories~\cite{yazdanshenas2017don}. We underscale $V_{CCBRAM}$ from nominal voltage to minimum operating voltage, $535mV$ (determined empirically).
\begin{figure}[ht]
  \centering
  \includegraphics[width=0.8\linewidth]{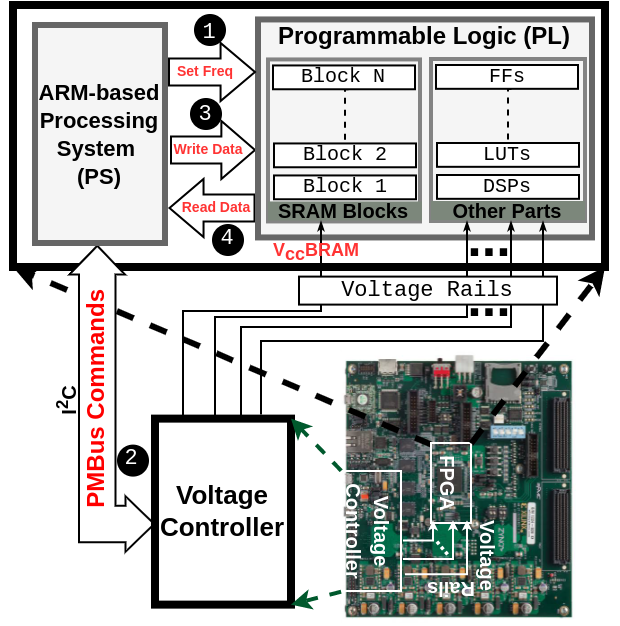}
  \caption{Overall Randomness Characterization Methodology, based on Supply Voltage underscaling in SRAMs}
  \label{fig:method}
\end{figure}
Our methodology is not only limited to our platform but also can be easily extended to various other FPGA-based platforms, given that their boards have the same independent voltage rail that the ZC702 board has. We expect not only the SRAM devices we test, but most of the SRAM devices to inherit random behavior when their supply voltage is underscaled, as prior works suggest~\cite{chen2005modeling, guo2009large,kim2011variation}.
Figure~\ref{fig:method} depicts the overall voltage underscaling methodology that we use for characterizing the randomness in SRAM blocks. 
\begin{algorithm}
\small
\caption{Voltage underscaling Randomness Testing Algorithm}
\label{alg:testalgo}
\begin{algorithmic}[1]
\REQUIRE $voltage, frequency, data\_pattern$
\STATE $set\_frequency(frequency)$
\STATE $reduce\_voltage(voltage)$
\STATE $write\_row(data\_pattern)\ into\ every\ row$
\FORALL{$row \in  SRAMs$}
\WHILE{$repeat < 1000$}
\STATE $value_{row} \leftarrow read\_row(row)$
\STATE $record(value_{row})$
\ENDWHILE
\ENDFOR
\end{algorithmic}
\end{algorithm}

We follow the general characterization methodology explained in Algorithm \ref{alg:testalgo} to study the randomness behavior of voltage underscaling-based faults on SRAMs. As shown in Algorithm \ref{alg:testalgo} and Figure \ref{fig:method}, \ding[1.4]{182} we first set the operating frequency (Line 1), and \ding[1.4]{183} reduce the supply voltage of SRAMs (Line 2). To reduce the voltage, corresponding PMBus commands are sent to the Voltage Controller from Processing System (PS). After adjusting the operating settings of SRAM blocks, \ding[1.4]{184} we write the corresponding data pattern on each row (Line 3). After writing, \ding[1.4]{185} we read every row 1000 times (Line 6) and record all 1000-bit bitstreams into off-chip memory~\textit{(i.e., DRAM)} for each row (Line 7). Then, we measure the entropy of bitstreams of each row using FPGA's PS side.

We use Shannon Entropy~\cite{shannon2001mathematical} mechanism to evaluate randomness in reduced-voltage SRAM rows. Shannon Entropy is calculated by Equation \ref{eq:shannon}.
\begin{equation}
\label{eq:shannon}
H(x) = -\sum_{i = 0}^{1}P(x_{i})\log_{2}P(x_{i})
\end{equation}
 $x$ denotes an arbitrary SRAM cell,
 $H(x)$ denotes the Shannon Entropy of the $x$, $P(x_{0})$ is the probability of logic-0 value, and $P(x_{1})$ denotes the probability of logic-1 value. We measure each cell's entropy and calculate their sum for every row.

We study the entropy of SRAM rows under two different parameters, voltage and data pattern. We perform these experiments at the nominal operating temperature and 200MHz as a operating frequency. After collecting raw data, we analyze the correlation between the supply voltage and operating frequency and their effects on entropy. Lastly, we study the effects of temperature on entropy for different voltage and temperature levels. We present average and maximum entropy for every 32-bit block on each parameter. Maximum entropy is the highest entropy of a 32-bit block across whole SRAM-based on-chip memories in an FPGA (total number of 280 SRAM chips). We use the average entropy term as the average entropy of all 32-bit blocks across all SRAM arrays in an FPGA.
\subsection{Voltage Level}
\label{subsec:volt_level}
To study how the supply voltage affects randomness we first start from the minimum operating voltage level, $535mV$, then the increase voltage level by $5mV$ in each iteration until maximum entropy is smaller than 1. We use the 0xFFFF as a data pattern and 200MHz as the operating frequency. 
\begin{figure}[ht]
\centering
\captionsetup[subfigure]{justification=centering}
\subfloat[Board-A]{{\includegraphics[width=0.5\linewidth]{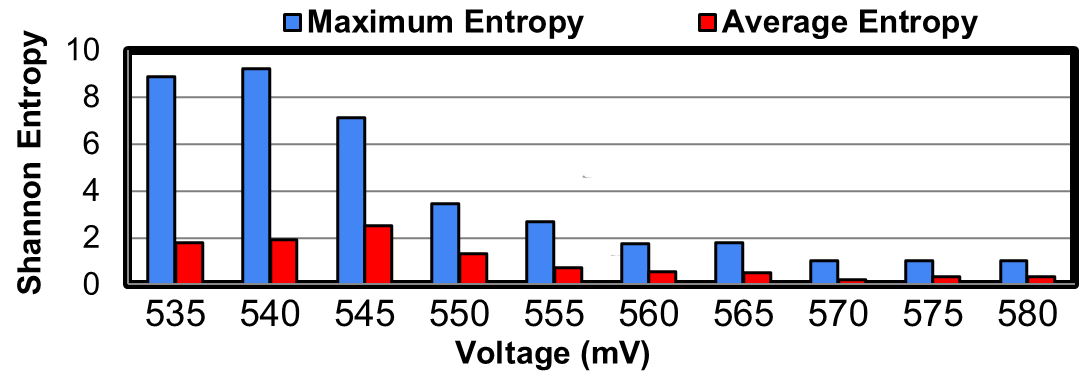} }\label{subfig:volt_board_a}}
\subfloat[Board-B]{{\includegraphics[width=0.5\linewidth]{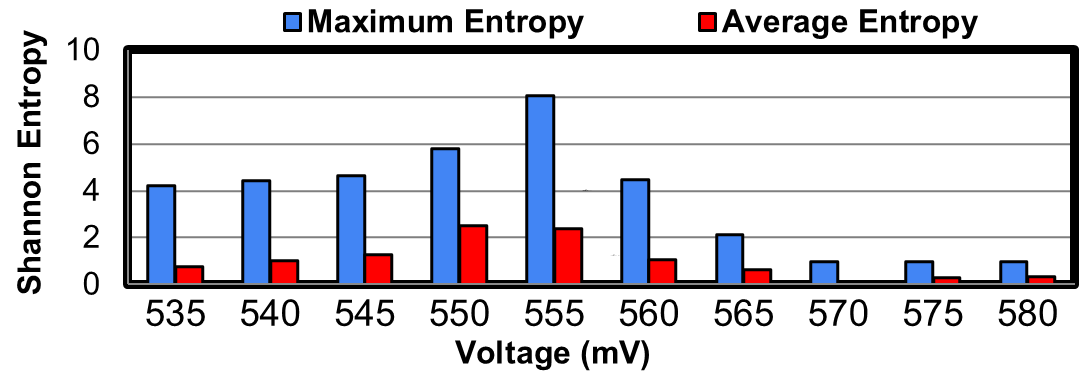} }\label{subfig:volt_board_b}}
\caption{Maximum and average entropy of 32-bit block for different voltage levels in each SRAM chip.}
\label{fig:volt_ent}
\end{figure}
Figure~\ref{fig:volt_ent} shows the average and maximum 32-bit block entropy for different voltage levels. The maximum entropy reaches its highest value above the minimum operating voltage level for both boards ($540mV$ for Board-A and $555mV$ for Board-B). We hypothesize that when the supply voltage is set as low as possible (e.g. 535 mV), voltage underscaling-based failures become deterministic as we always observe faults (100\% probability). Accordingly, setting slightly higher voltage levels decreases the number of cells that fails at 100\% probability and increase the number of SRAM cells that exhibit a access failure rate of ~50\%, resulting in higher entropy. 

\subsection{Data Pattern}\label{subsec:dp_sec}
\begin{figure}[ht]
\captionsetup[subfigure]{justification=centering}
\subfloat[Board-A]{{\includegraphics[width=0.5\linewidth]{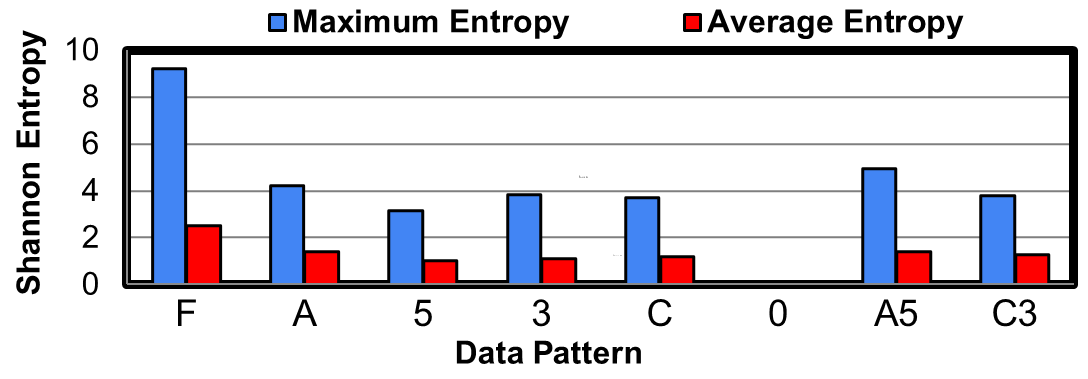} }\label{subfig:dp_board_a}}
\subfloat[Board-B]{{\includegraphics[width=0.5\linewidth]{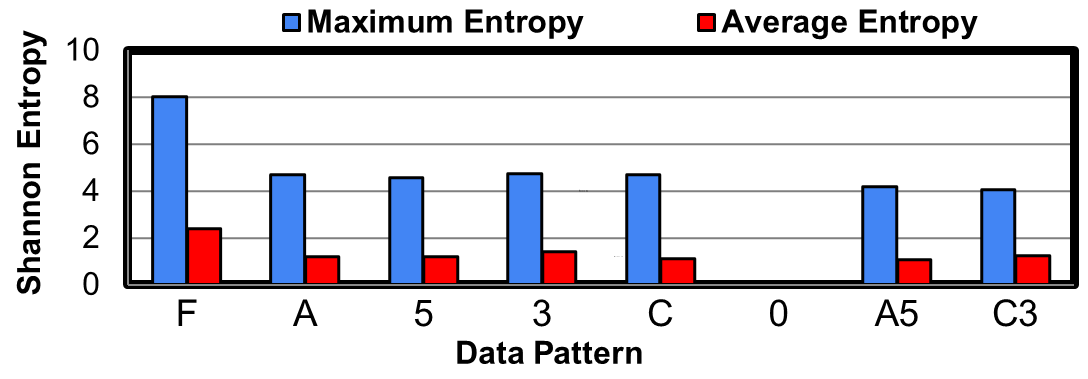} }\label{subfig:dp_board_b}}
\caption{Maximum and average 32-bit block entropy for different data patterns in each SRAM chip.}
\label{fig:dp_ent}
\end{figure}

To study the effects of data patterns on entropy, we test eight different data patterns. We set the voltage and frequency to the level where the maximum entropy is highest for both boards. We analyze the impact of data patterns on entropy by two approaches. First, in the traditional method, values are written to each row with the same data pattern. For this first method we use six different data patterns; 0xFFFF, 0xAAAA, 0x5555, 0x0000, 0x3333, and 0xCCCC. The second approach is based on writing different values in two consecutive rows. In the same column, bit-0 is written to one and bit-1 to the other to understand if consecutive two rows affect each other. We use 0xAAAA with 0x5555 and 0xCCCC with 0x3333 to evaluate this approach. In Figure~\ref{fig:dp_ent}, we refer the values of first approach with their first 4-bit value, such as F stands for 0xFFFF, A stands for 0xAAAA. Also, for the second approach we use A5 for 0xAAAA with 0x5555 and C3 for 0xCCCC with 0x3333. Figure~\ref{fig:dp_ent} depicts the average and maximum entropy across the data patterns.
We make three key observations from Figure~\ref{fig:dp_ent}. First, the data pattern behavior of these boards is identical in contrast to the previous two experiments. The 0x0000 data pattern has the lowest value for maximum and average entropy. In addition, the highest maximum entropy is observed with a 0xFFFF value. This is because bitflips mostly happen in cells that have logic-1 value. This is in line with the results of prior work~\cite{salami2018comprehensive}. Second, other values of the first approach have nearly the same values in terms of maximum and average entropy which contain same the number of logic-1 in their data. Third,consecutive rows do not affect each other.

\subsection{Voltage and Frequency Correlation}\label{subsec:vf_sec}
In Section~\ref{subsec:volt_level}, we only analyze 200MHz operating frequency behavior in different voltage levels and show that in Figure~\ref{fig:volt_ent}, the highest maximum entropy is above the minimum operating voltage. Therefore, we study the randomness behavior of different frequencies at different voltage levels. 
    
To analyze, we choose 20MHz, 60MHz, 100MHz, 160MHz, and 200Mhz as frequency levels and 0xFFFF as the data pattern. Also, the range of voltage level is from the minimum operating voltage (i.e., $535mV$) to $580mV$ where the maximum entropy is saturated and does not change afterward, as can also be seen in Figure~\ref{fig:volt_ent}.

    \begin{figure}[ht]
    \centering
    \captionsetup[subfigure]{justification=centering}
    \subfloat[Board-A]{{\includegraphics[width=\linewidth]{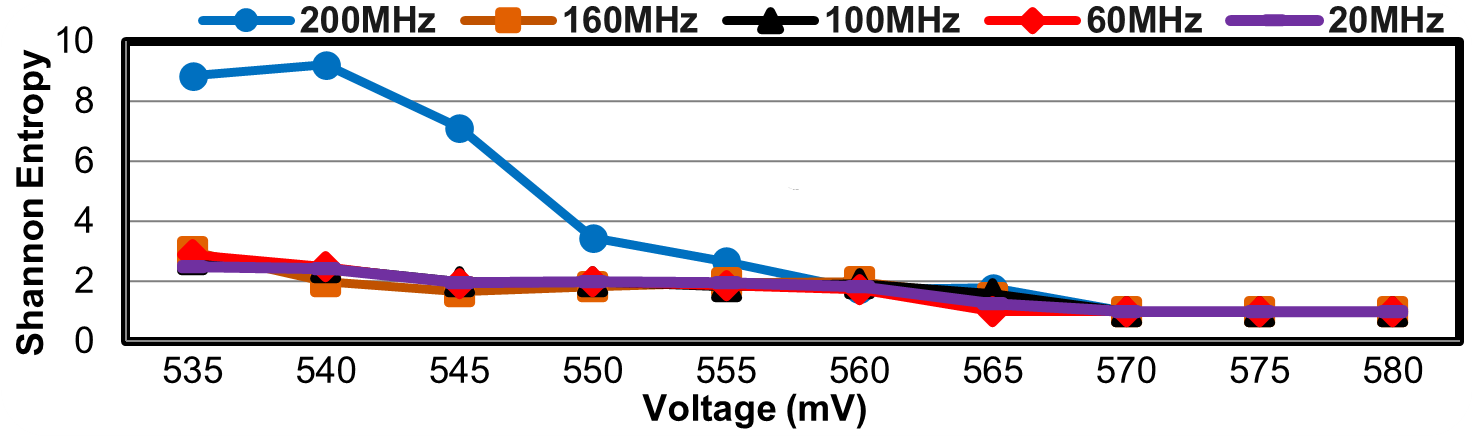} }\label{subfig:vf_board_a}}
    
    \subfloat[Board-B]{{\includegraphics[width=\linewidth]{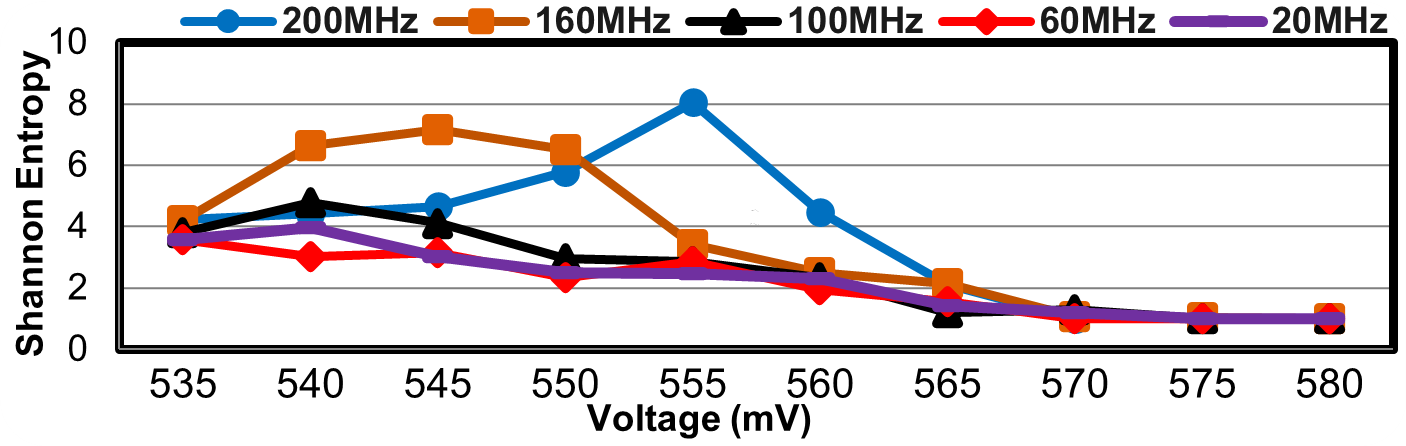} }\label{subfig:vf_board_b}}
    \\
    \caption{Maximum 32-bit block entropy across different sets of operating voltage and frequency in each SRAM chip.}
    \label{fig:vf_ent}
    \end{figure}
  Figure~\ref{fig:vf_ent} depicts the average and maximum entropy of different sets of voltage and frequency parameters for both SRAM chips. Similar to the Figure~\ref{fig:volt_ent} and Figure~\ref{fig:freq_ent}, we observe that \textit{(i)} the highest maximum entropy is achieved at \emph{different} operating voltage levels for both boards and frequency levels, such as in Figure~\ref{subfig:vf_board_b}, \textit{(ii)} the voltage level that achieves peak value of the maximum entropy is not same for every frequency and \textit{(iii)} from \ref{subfig:vf_board_a} and \ref{subfig:vf_board_b} in a various different set of voltage and frequency parameters 200MHz has the highest maximum entropy, 9.21 and 8.04 for Board-A and Board-B, respectively.
  
  When the voltage is set as low (e.g., 535 mV) and the frequency is set as high (200 MHz) as possible the probability of access failure for the largest majority of SRAM cells reaches 100\%. We observe that the number of SRAM cells that exhibit a access failure rate of ~50\% is maximized at 555 mV. When we increase the voltage beyond 555 mV, the number of SRAM cells that fail with a 0\% probability increases as the number of SRAM cells that fail with a 50\% probability decreases. Hence, we observe a non-monotonic behavior seen in Figure~\ref{subfig:vf_board_b}.
  
 \subsection{Temperature}
 \label{subsec:temp}
\begin{figure}[ht]
  \centering
  \includegraphics[width=1\linewidth]{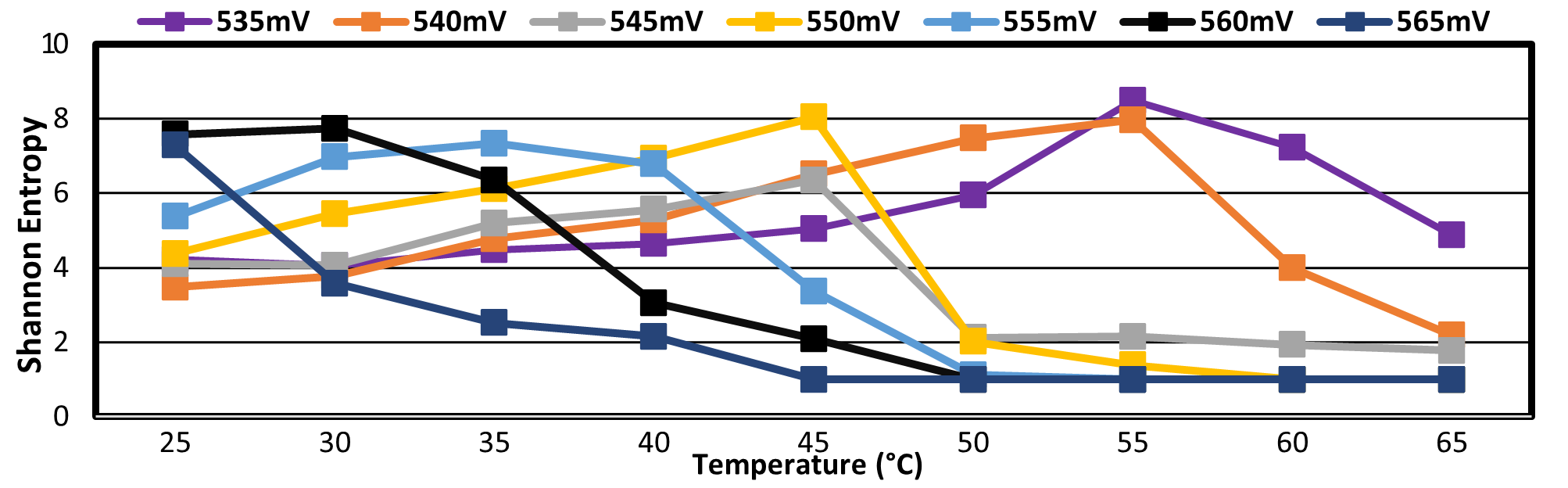}
  \caption{Maximum 32-bit block entropy across different sets of operating voltage and temperature in Board-B}
  \label{fig:temp}
\end{figure}

 We study the effect of the enviromental temperature on entropy. To perform this experiment, we monitor the on-board live temperature of an FPGA board using PMBus interface. We set the frequency to 200MHz for both boards and use 0xFFFF as a data pattern. We analyze the entropy under different pairs of temperature and voltage levels ranging from $25^\circ C$ to $65^\circ C$ and from 535mV to 565mV, respectively. Figure~\ref{fig:temp} depicts the highest 32-bit block entropy across different pairs of operating voltage and temperature for Board-B. It should be noted that, we also observe similar behavior for Board-A. We highlighted three voltage levels, 565mV as the highest tested voltage level, 535mV as the lowest voltage level, and 550mV which achieves the highest entropy at the nominal temperature ($45^\circ C$). 

We make three key observations from Figure~\ref{fig:temp}, 1) the highest maximum entropy is achieved at different voltage level for most temperature levels (e.g., 535mV at $65^\circ C$, 560mV at $25^\circ C$), 2) at lower temperatures higher voltage levels achieves the highest entropy (e.g. at $25^\circ C$ the highest entropy is produced from 565mV) and at higher temperatures lower voltage levels achieves the highest entropy(e.g. at $>55^\circ C$ the highest entropy is produced from 535mV), and 3) at each temperature level reduced-voltage SRAM exhibit randomness and at least has one voltage level that can produce $>7$ entropy. We conclude that since the SRAM rows' entropy is affected by temperature, TuRaN needs to take temperature changes into account while generating true random numbers. We discuss the robustness and reliability aspects of TuRaN in Section~\ref{subsec:sysint-dis}. 

\subsection{Discussion}
\label{characterization:discussion}
The characterization phase in TuRAN is a one-time and low-cost process that identifies the reduced-voltage SRAM cells that can be used as a source of entropy. In this section, we first discuss the impact of aging and process variation on the entropy of SRAM cells. And lastly, we discuss and hypothesize why we observe randomness when we underscale the supply voltage of SRAM blocks.

\noindent
\textbf{Time Dependence:} To ensure that SRAM aging does not adversely affect the entropy of reduced-voltage SRAM cells, we empirically evaluate the aging of our characterization results and repeat all experiments one year later. We successfully reproduce the same results. Therefore, we expect that the randomness characterization results are valid at least for a year.

\noindent
\textbf{Process Variation:} Prior work~\cite{salami2018comprehensive} shows that the voltage guard-band and the minimum operating voltage level vary across different SRAM chips (similar observations hold for other memory technologies~\cite{larimi2021understanding,chang2017understanding}) due to process variation. We perform randomness characterization on two FPGA boards and observe different behavior in maximum and average entropy under identical operating conditions. We conclude that randomness behavior can change across different SRAM devices. Thus, randomness characterization has to be performed once for every SRAM chip.

\section{TuRaN: An SRAM-based TRNG}\label{label:turan}
Based on our randomness analysis of undervolting failures in SRAM cells, we propose TuRaN, a new SRAM-based TRNG that performs voltage underscaling in SRAM blocks, and processes the resulting faults using a cryptographic hash function, SHA-256. TuRaN leverages the observation that when the voltage is reduced below the safe voltage margin, SRAM cells fail at sensing operation indeterminately, and this non-deterministic failures can be used as a source of entropy. TuRaN consists of three steps: 1) setting the operating parameters (frequency, supply voltage, and data pattern) using one-time characterization phase, 2) reading previously-characterized rows that have the highest entropy, and 3) post-processing each block by performing the SHA-256 hash function to obtain a high-quality, 256-bit true random number.
\subsection{TuRaN Evaluation}\label{sec:eval}
We evaluate TuRaN on off-the-shelf SRAM chips embedded in FPGA boards. We perform our evaluation on two identical samples of Xilinx ZC702 FPGA boards~\cite{zc702board}. We use this platform to evaluate TuRaN since it enables us 1) to easily manipulate both frequency and supply voltage of SRAM chips, 2) to perform fast empirical experimentation as the post-processing hardware can be implemented into this platform 3) to monitor energy consumption via a voltage regulator/power controller. We evaluate TuRaN in four categories. First, we evaluate the quality of the generated random numbers using the standard NIST STS randomness tests~\cite{bassham2010sp}. Second, we analyze the throughput of TuRaN for different frequency levels. Third, we evaluate the energy consumption of TuRaN. Fourth, we evaluate TuRaN's true random number generation latency. We choose the supply voltage for each frequency level, based on the voltage level at which the highest entropy is observed (e.g. for Board-A at 200MHz, $540mV$.). 
We show that TuRaN successfully generates high-quality true random numbers with high-throughput, high energy efficiency, and low-latency. 

\subsubsection{Quality}
\label{subsec:quality}
To evaluate the quality of random numbers generated by TuRaN, we extract random bitstreams from both FPGAs. We generate a 1Gbit random bitstream and partition it into 1024 sequences each with the length of 1Mb ($2^{20} bit$). We test these 1024 sequences using the NIST Statistical Test Suite (STS)~\cite{bassham2010sp} tests. NIST STS is used to evaluate randomness by formulating several statistical tests. Each test has a p-value that indicates the status of the null hypothesis of the test. If the p-value is greater than the significance level \textit{i.e, $\alpha$}, the null hypothesis of the test holds (\textit{i.e.,} the sequences are truly random).

\begin{table}[h]
\centering
\small
\caption{NIST STS Randomness Test Results for TuRaN}
\label{tab:nist_table}
\resizebox{\linewidth}{!}{%
\begin{tabular}{@{}ccc@{}}
\toprule
\multicolumn{1}{c}{\textbf{NIST Test Name}} & \multicolumn{1}{c}{\textbf{p-value ($\alpha=0.01$)}} & \multicolumn{1}{c}{\textbf{Test Status}} \\ \midrule
Frequency                 & 0.42649 & PASS \\
Block Frequency           & 0.24730 & PASS \\
Cumulative Sums           & 0.38451 & PASS \\
Runs                      & 0.63712 & PASS \\
Longest Run               & 0.09818 & PASS \\
Rank                      & 0.55003 & PASS \\
DFT                       & 0.07785 & PASS \\
Non-Overlapping Template  & 0.51272 & PASS \\
Overlapping Template      & 0.67787 & PASS \\
Universal                 & 0.84941 & PASS \\
Approximate Entropy       & 0.28524 & PASS \\
Random-Excursions         & 0.67243 & PASS \\
Random-Excursions Variant & 0.52986 & PASS \\
Serial                    & 0.58120 & PASS \\
Linear Complexity         & 0.01383 & PASS \\ \bottomrule
\end{tabular}%
}
\end{table}

Table~\ref{tab:nist_table} shows the average results of 1024 1Mbit sequences in terms of p-value across the all 15 tests for randomness where the $\alpha = 0.01$.Our results show that $99.02\%$ of the 1Mbit sequences (1024 in total) pass each NIST test. This percentage is in the acceptable range ($>98.84\%$) determined by NIST for STS tests~\cite{bassham2010sp}.This indicates that TuRaN generates high-quality truly random numbers. The required number of reads to generate true random numbers differs for various operating frequency levels. This is because lower frequency levels tend to have lower entropy as we observe in Figure~\ref{fig:vf_ent}. Accordingly, to have a thorough analysis, we perform our evaluation on five different frequency levels (20MHz, 60MHz, 100MHz, 160MHz, and 200MHz) and find that the number of reads is (85, 79, 66, 50, 32), respectively associated with the frequency levels aforementioned.

TuRaN generates true random bitstreams even without any post-processing (e.g., without SHA-256). We conduct a new experiment and find 37 true random SRAM cells that can be used without any post-processing. We repeatedly perform TuRaN on each true random SRAM cell to generate a 1 Mbit true random bitstream. These bitstreams pass all the NIST tests. We conclude that TuRaN leverages unpredictable random physical phenomena to generate random values in SRAM sense amplifiers.

\subsubsection{Throughput}
\label{sec:throughput}
To evaluate the throughput of TuRaN, we first determine the required number of read operations to the SRAM row with the highest entropy to accumulate 256-bit of entropy. Then, we calculate the impact of the post-processing (SHA-256) step on latency and throughput.

\begin{figure}[!h]
\centering
\captionsetup[subfigure]{justification=centering}
\subfloat[Throughput]{    \includegraphics[width=0.5\linewidth]{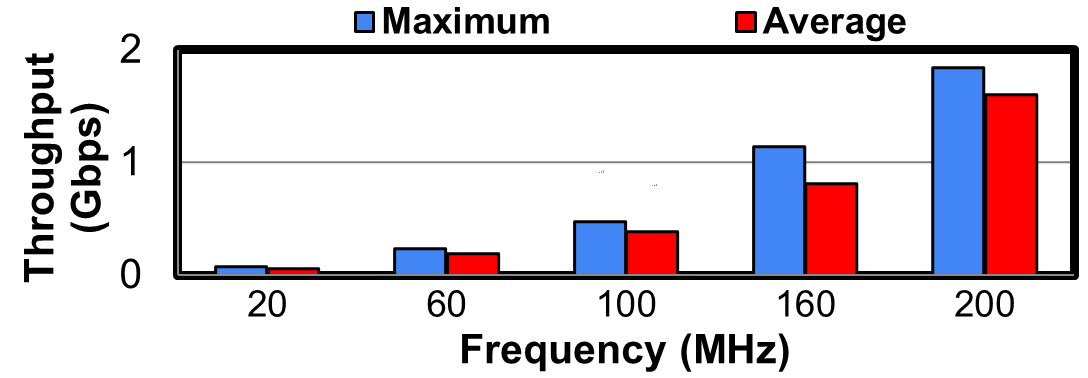}\label{fig:throughput}}
\subfloat[Energy Consumption]{    \includegraphics[width=0.5\linewidth]{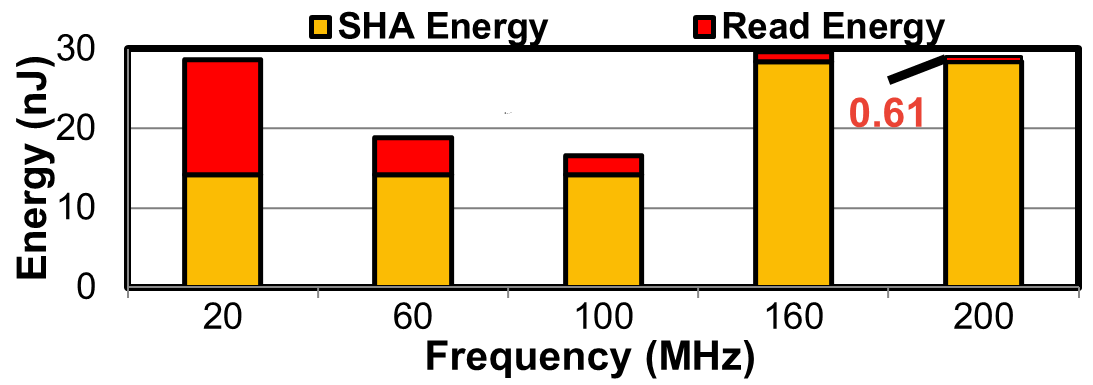}\label{fig:energy}}
\caption{Maximum and average throughput and energy consumption of TuRaN at different operating frequency levels}
\label{fig:freq_ent}
\end{figure}

Figure~\ref{fig:throughput} shows the average and the maximum throughput achieved at five different frequency levels, characterized in Section \ref{sec:characterization} for two sets of SRAM blocks. We observe that increasing the frequency \emph{exponentially} increases the throughput. For instance, at 200MHz maximum throughput is $1.812Gbps$, $25.8x$ higher than the observed throughput at 20MHz. This is because higher frequency levels achieve higher entropy which decreases the required amount of read operations to accumulate 256-bit of entropy.

\subsubsection{Energy}
We evaluate the energy consumption of TuRaN in two steps: 1) we monitor the energy consumption of read operation using PMBus to obtain voltage and current values, 2) we calculate the energy of the SHA-256 hash function to generate 256-bit true random numbers. In the first step, we monitor the current and voltage rail of SRAM blocks to obtain the energy consumption of read operations. Second, we derive the power and throughput results of SHA-256 from recent work~\cite{kammoun2020fpga}, as they use the same FPGA board and propose a design that does not use any SRAM blocks to perform SHA-256 hash operation. This study reports that one SHA-256 hardware achieves $917Mbps$ while consuming $0.1W$.

Figure~\ref{fig:energy} shows the energy consumption achieved at five different frequency levels to generate 256-bit true random numbers. We make three key observations from Figure \ref{fig:energy}: 1) The SHA-256 accelerator dominates the total energy consumption in each frequency level except 20MHz. 2) Although the power consumption of read operations decreases in lower frequencies, the energy consumption of total read operations is higher in lower frequency levels as the number of read operations and latency increase when SRAM operates at a lower frequency. 3) After 100MHz, the energy consumption of the SHA-256 accelerator increases as the number of SHA-256 accelerators is doubled to achieve the maximum throughput in 160MHz ($1.144 Gbps$), and 200MHz ($1.812Gbps$). Our results show that TuRaN consumes $0.11nJ$ to generate a one-bit true random number.

\subsubsection{Latency}
The latency of TuRaN is directly related to 1) the setup time of PMBus to manipulate voltage rails, 2) the execution time of the voltage underscaling command 3) SRAM access latency (including write and read operations) and 4) the setup time of the post-processing function.

We measure the latency of 1st, 2nd, and 3rd operations by using the ARM-based Processing System (PS) of ZC702. For the fourth operation, we use the prior work's observation~\cite{kammoun2020fpga}. In the setup time of PMBus which takes $228.3\mu s$, the system is initialized with related configuration parameters and registers to underscale the supply voltage of SRAMs. After initializing the system, we send the undervolting command to reduce voltage with the latency of $49.7\mu s$. In the third operation, we access a row in the reduced-voltage SRAM to obtain input bitstreams for SHA-256 which takes $320ns$ at 200MHz. In the last step of generating random numbers, we perform the SHA-256 operation to generate 256-bit true random numbers in $142.2ns$. At 200MHz, we obtain $278.46\mu s$ latency.  
Since all steps except the third step are independent of any operating conditions, the difference in total latency between different frequencies is determined by this step. We observe the lowest latency at 200MHz, $278.46\mu s$, and the highest latency at 20MHz, $282.39\mu s$.

\subsection{Impact of Environmental Factors on TuRaN}
We study the effect of temperature and aging on entropy in Section~\ref{subsec:temp} and Section~\ref{characterization:discussion}, respectively. In this section, we analyze the impact of temperature and time dependence on TuRaN in terms of quality, throughput, energy consumption, and latency.

\noindent
\textbf{Time Dependence:} We monitor the throughput, energy consumption, and latency of TuRaN over the course of one year and three months. We observe that one year and three months after the initial evaluation results, TuRaN generates true random numbers with the same throughput, energy consumption, and latency for each SRAM device. We believe that since the entropy value of the source is directly related to the quality of random numbers and the entropy is not affected by aging for at least one year, we obtain the same results.

\begin{figure}[!h]
\centering

\captionsetup[subfigure]{justification=centering}
\subfloat[Throughput]{    \includegraphics[width=0.33\linewidth]{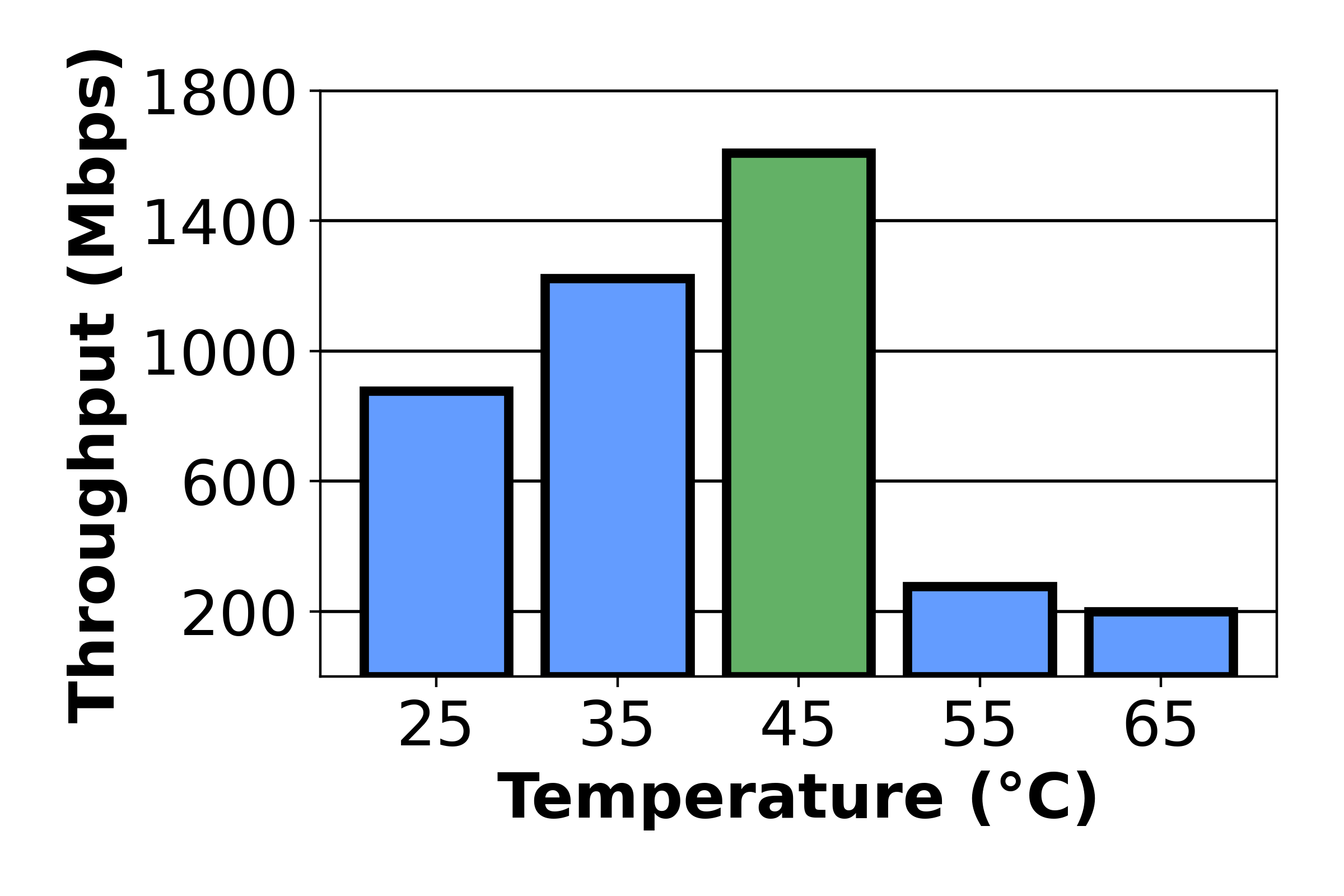}\label{fig:temp_throughput}}
\subfloat[Energy Consumption]{    \includegraphics[width=0.33\linewidth]{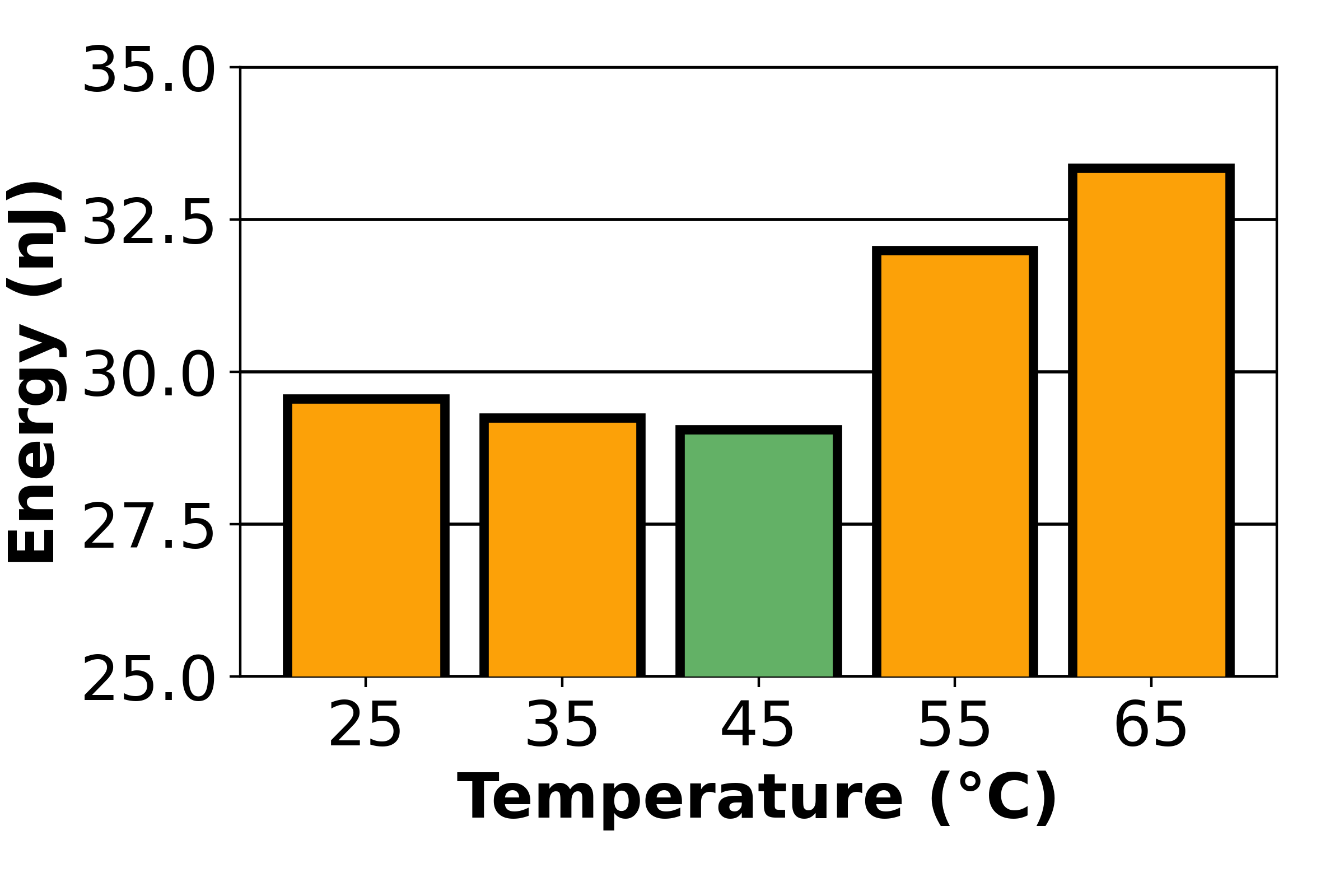}\label{fig:temp_energy}}
\subfloat[Latency]{
\includegraphics[width=0.33\linewidth]{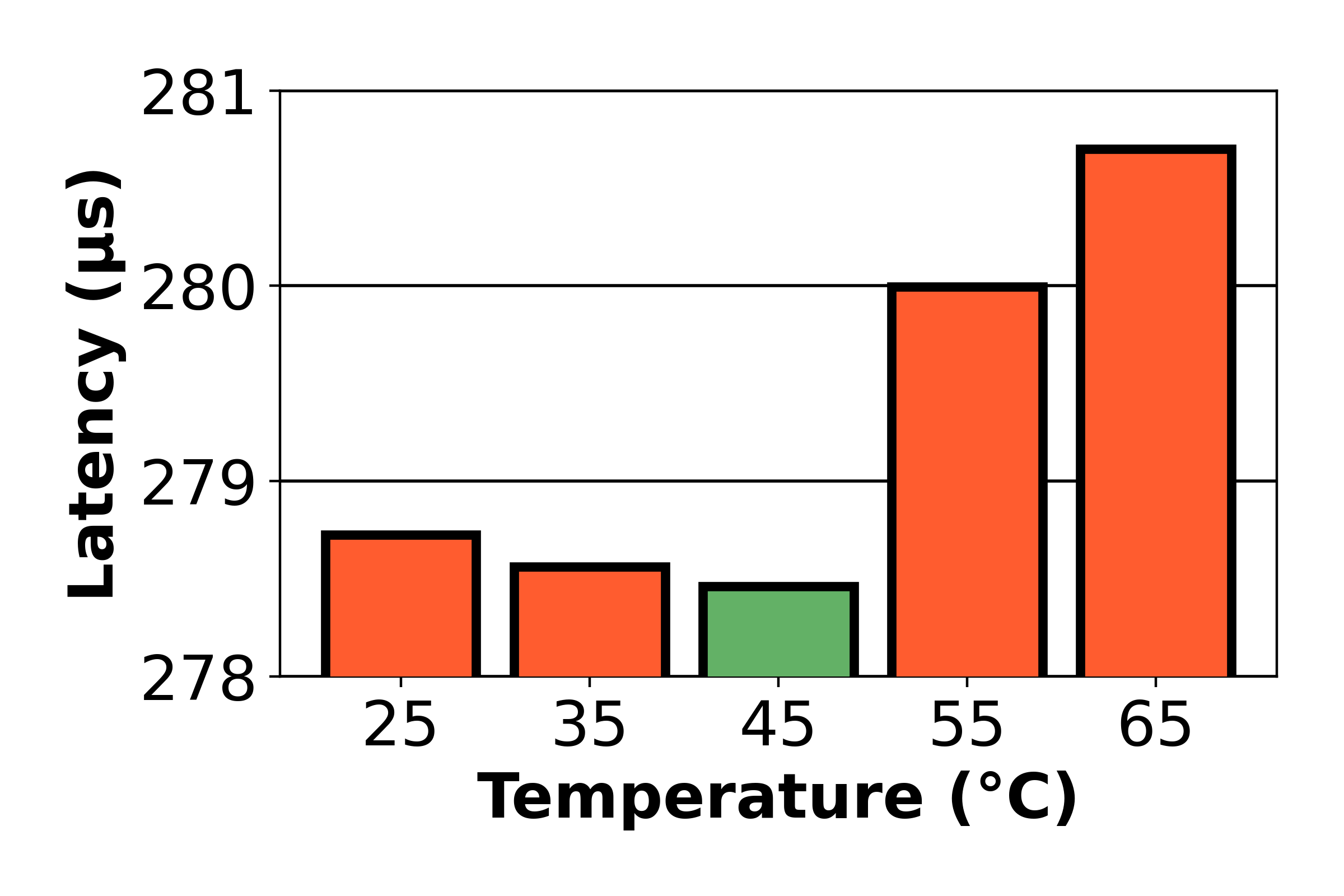}\label{fig:temp_latency}}
\caption{The impact of temperature on evaluation parameters}
\label{fig:temp_eval}
\end{figure}

\noindent
\textbf{Temperature:} We analyze all the evaluation parameters (quality, throughput, energy consumption, and latency) of TuRaN for different temperature levels ranging from $25^\circ C$ to $65^\circ C$ at 200MHz, 550mV for Board-B. Figure~\ref{fig:temp_eval} shows the average evaluation parameters for different temperature levels (green bar depicts the nominal temperature). We observe that 1) TuRaN reliably generates true random numbers regardless of the temperature level and 2) as the entropy is affected by temperature changes (see Fig.~\ref{fig:temp}), TuRaN is also affected by temperature and has different throughput, energy consumption, and latency for each temperature level.

\sloppy
\subsection{Comparison with the state-of-the-art SRAM-based TRNGs}
\label{subsec:comparison}
To our knowledge, this is the first study that uses undervolting-based faults in SRAMs as a source of entropy. We compare TuRaN with state-of-the-art SRAM-based TRNGs in terms of continuous operation, peak throughput, energy consumption per bit, and 256-bit latency. Table~\ref{tab:comparison} shows a summary and comparison of TuRaN and previous state-of-the-art SRAM-based TRNGs~\cite{li2015pufkey,zhang2020improved}. We demonstrate that TuRaN meets all properties that SRAM-based TRNGs must have (Section \ref{sec:motivation}) and is superior to any prior works in all comparison points, $2.26x$, $5.39x$, $5.09x$ in throughput, energy efficiency, and latency, respectively.   

Unfortunately, the state-of-the-art SRAM-based TRNG proposals~\cite{zhang2020improved,li2015pufkey} do not report their total latency. Since the prior work~\cite{zhang2020improved} empirically evaluates the latency of the power-down period, we use their value, $250ms$ as a power cycle latency for works that do not mention their power cycle latency. Also, they do not mention the operating frequency and the supply voltage of SRAM. To evaluate the remaining parameters (latency, and energy consumption) we optimistically assume that for each prior work SRAM operates at 200MHz and the supply voltage is $1V$, which is the nominal voltage level of our evaluation. We evaluate the energy consumption of prior works using PMBus on our platform by calculating their number of read operations to obtain 256-bit entropy\footnote{We assume that each SRAM cell of prior works has \emph{1-bit} (totally random) entropy which is the ideal situation, also $3.47x$ higher than TuRaN.}. 

\textbf{Zhang+~\cite{zhang2020improved}:} Zhang et al. propose an SRAM-based TRNG that improves the TRNG performance by utilizing ionization irradiation on SRAM.The authors implement SHA-256 hardware that operates at 200MHz on the ZC702 FPGA board (same as TuRaN) to post-process. Also, Zhang et al. report their throughput as $178Mbps$. However, they do not mention any energy consumption of their TRNG. Based on these optimistic parameters, Zhang+'s TRNG consumes $0.56nJ$ per true random bit. To evaluate the latency of the proposed TRNG, we consider the latency of their improved power cycle (1.5ms latency), SRAM read access latency, and the latency of the post-process function. The latency of Zhang+'s SRAM-based TRNG is $1.501ms$.

\textbf{PUFKEY~\cite{li2015pufkey}:} PUFKEY generates true random numbers by using two different physically unclonable functions (PUFs). The first step is to obtain true random seeds using a conditional algorithm (u-Quark). Second, true random seeds is used as an input for a hardware RNG (HRNG) to generate true random numbers. The authors report that PUFKEY achieves $803Mbps$ throughput. From their observation, u-Quark needs 0.0255 seconds. To achieve $803Mbps$, we assume that authors implement 52.4 u-Quark blocks that have a latency of $5.1s$. The energy consumption of the second step, \textit{i.e.,} NDRNG, a specialized hardware is not reported. Thus, we cannot calculate the energy consumption of PUFKEY. The latency of NDRNG is reported as $159.22ns$. Based on these calculations and observations total latency of PUFKEY is $5.35s$.
\begin{table}[ht]
\small
\caption{TuRaN vs prior SRAM-based TRNGs}
\resizebox{\linewidth}{!}{%
\begin{tabular}{lcccc}
\toprule
\multicolumn{1}{c}{\textbf{Proposal}} &
  \textbf{\begin{tabular}[c]{@{}c@{}}Continuous\\ Operation\end{tabular}} &
  \textbf{\begin{tabular}[c]{@{}c@{}}Peak \\ Throughput\end{tabular}} &
  \textbf{\begin{tabular}[c]{@{}c@{}}Energy \\ Consumption\end{tabular}} &
  \textbf{\begin{tabular}[c]{@{}c@{}}256-bit \\ Latency\end{tabular}} \\ \midrule
Zhang+\cite{zhang2020improved} & \ding{55}  & $178 Mbps$   & $0.56nJ/bit$  & $1.501ms$  \\
PUFKEY~\cite{li2015pufkey} & \ding{55} & $803 Mbps$   & $N/A$        & $5.35s$    \\
\textbf{TuRaN}  & \ding{51} & \textbf{$1.812 Gbps$} & \textbf{$0.11nJ/bit$} & \textbf{$278.46\mu s$} \\ \bottomrule
\end{tabular}
}
\label{tab:comparison}
\end{table}
\section{System Integration}
\label{sec:sysint}
TuRaN can be integrated into a computing system to generate true random numbers that are required by a wide variety of applications as discussed in Section~\ref{sec:motivation}.
Modern computing systems already employ multiple SRAM-based memory structures such as caches, branch predictor tables, translation lookaside buffers, and coherence directories. Existing SRAM-based memory structures can be used as a basis for integrating TuRaN at low hardware cost (e.g., complexity and area), whereas the variety of SRAM-based memory structures presents the system designer with multiple options with different system integration tradeoffs for TuRaN. For example, branch predictor tables are tightly integrated with the processor, thus random numbers can be retrieved quickly from the predictor tables to the processor. However, branch history tables typically have small row sizes, thus the maximum entropy that can be generated by accessing a single predictor table row is small, which constrains the TRNG throughput that can be obtained using TuRaN.

We discuss how TuRaN can be integrated into a modern system at low hardware cost to generate true random numbers at high throughput. We implement TuRaN in processor caches, because this design strikes a balance between TRNG latency (i.e., short distance from the processor core), and TRNG throughput (i.e., high entropy from large SRAM rows).

\subsection{Mechanism}
TuRaN encompasses two steps to generate true random numbers: 1) reading reduced-voltage rows from SRAM until obtaining a bitstream with 256-bit entropy, and 2) sending the output of the first step to the SHA-256 function. To enable the first step in modern computing systems,
TuRaN requires control over the supply voltage of rows in the cache (i.e., the cache line supply voltage) in order not to corrupt other lines' data when undervolting is performed. We implement TuRaN on top of the Drowsy Cache~\cite{flautner2002drowsy}, a low-cost substrate that allows fine-grained control over the supply voltage of cache lines. Drowsy Cache allows us to scale an arbitrary cache line’s voltage independently from others. Therefore, TuRaN does not undervolt the whole cache but only undervolts the cache line with the highest entropy after the initial characterization.
For the second step, TuRaN performs post-processing with SHA-256 cryptographic hash function using the CPU to avoid additional area overhead on commodity systems.

\noindent
\textbf{Drowsy Cache~\cite{flautner2002drowsy}.} The Drowsy Cache adds a drowsy bit, a word-line gating circuit and a voltage controller to every cache line to control the voltage of cache lines. When a cache line is accessed, the cache controller reads its drowsy bit to determine the level of the supply voltage. The supply voltage of the cache line is scaled by the voltage controller, which switches the voltage of the cache line between the nominal and low (i.e., drowsy) supply voltages depending on the drowsy bit. When a drowsy cache line is accessed, the supply voltage of the cache line is switched to the nominal voltage. 
The cache controller periodically put the cache line into drowsy mode to save energy. Implementing the Drowsy Cache induces a small area overhead of $<3\%$~\cite{flautner2002drowsy}.

\noindent
\textbf{TuRaN on the Drowsy Cache.} For TuRaN's integration, we propose two changes to the Drowsy Cache design to generate random numbers by enabling access failures in cache lines. First, TuRaN removes the word-line gating circuit to exploit the corrupted data as a source of entropy. Second, to avoid corrupting valid data in the cache, TuRaN does not put cache lines into drowsy mode periodically.

\noindent
\textbf{Cache Line Entropy Characterization.} To generate true random numbers, TuRaN first characterizes (as described in Section~\ref{sec:characterization-methodology}) the cache under reduced voltage to find the cache line with the highest entropy. Then, TuRaN stores the observed entropy of the cache line in a register, $r_{entropy}$, in the cache controller. 

\noindent
\textbf{Generating Random Bitstreams.} TuRaN generates true random numbers in the cache in four steps. First, TuRaN writes an all-ones data pattern (i.e., all cells in the row is filled with logical-1) to the highest entropy cache block. Second, TuRaN switches the mode of the cache line to drowsy mode to apply undervolting. Third, TuRaN reads the cache block in the drowsy mode to retrieve a $64B$ bitstream. Fourth, TuRaN switches the mode of the cache line back to normal mode (i.e., apply nominal operating voltage). Each step takes one cycle to execute. In total, it takes four cycles to obtain a $64B$ bitstream with $r_{entropy}$ bits of entropy. Based on our observation (Section~\ref{sec:characterization}) that a 32-bit SRAM row can contain more than 8 bits of entropy, we assume that the cache line entropy characterization step can identify a cache line with at least 128 bits of entropy (because a cache line is 16$\times{}$ larger than a 32-bit SRAM row). Thus, we use a $128B$ $r_{random}$ in our evaluation to accumulate 256 bits of entropy with two cache line accesses. TuRaN stores the bitstream in the $r_{random}$ register in the cache controller. TuRaN repeatedly performs these four steps until $r_{random}$ contains 256 bits of entropy.

\noindent
\textbf{SHA-256 Operation in CPU.} To post-process the obtained bitstream ($r_{random}$), TuRaN performs the SHA-256 cryptographic hash function. TuRaN uses the CPU to perform SHA-256 because (i) contemporary CPUs are equipped with special SHA-1 and SHA-256 instructions~\cite{guilford2012fast}, this enables TuRaN to perform SHA-256 with a throughput of 27.984 Gbps~\cite{sha256-simd}, and (ii) it does not require a dedicated SHA-256 hardware, thus it does not cause an additional area and energy overhead.

\subsection{Evaluation}
\begin{table}[h]
\centering

\resizebox{\linewidth}{!}{%
\begin{tabular}{@{}ll@{}}
\toprule
\multicolumn{1}{l}{\textbf{Parameter}} & \multicolumn{1}{l}{\textbf{Value}} \\ \midrule
\textbf{Processor Type}      & Out-of-order x86 CPU \\
\textbf{Processor Base Frequency} & 3.6GHz           \\
\textbf{L1 Data Cache (Latency)}            & 32KiB, 8-way, LRU, Set-associative (2 cycles)       \\ 
\textbf{L2 Cache (Latency)}            & 256KiB, 4-way (12 cycles)      \\
\textbf{L3 Cache (Latency)}            & 2MB, 16-way  (44 cycles)        \\
\textbf{DRAM Memory}            & DDR4, 2400MHz, 8GB, 2 channels     \\ \bottomrule
\end{tabular}%
}
\caption{gem5 simulation parameters}
\label{tab:gem5eval}
\end{table}

To estimate TuRaN's random number generation benefits in modern systems, we perform simulations using gem5~\cite{binkert2011gem5}. We run single-core applications from SPEC2006 benchmark suite on a simulated system. The characteristics of the simulated system can be found in Table \ref{tab:gem5eval}. As the highest-entropy yielding cache line can change across different chips, we simulate TuRaN using different cache lines as entropy sources across a way of the cache. We run the applications for each cache line and for every run, we select a different cache line and always evict that line to generate random numbers. We analyze the idle cycles of the L1 data cache and the L2 cache, then inject TuRaN mechanism commands into these idle intervals. 
Since the simulated system runs at 3.6 GHz clock frequency, we estimate that the cache line with the highest entropy have our 200MHz's entropy results (Section~\ref{sec:characterization}) as it is the closest evaluated frequency level to 3.6GHz.

\subsection{Results}
\begin{figure}[ht]
  \centering
  \includegraphics[width=\linewidth]{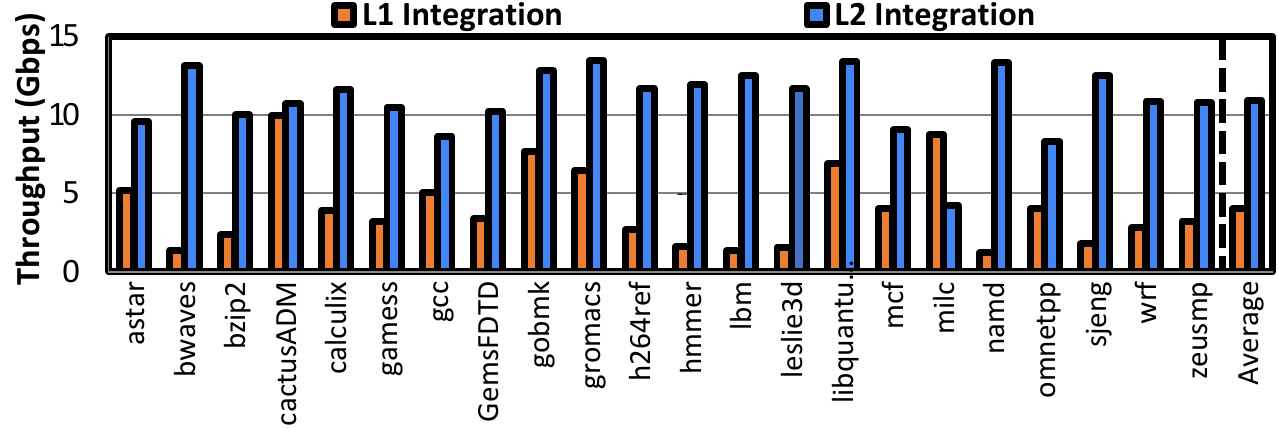}
  \caption{Average throughput of two integration scenarios of TuRaN in modern systems for SPEC2006 workloads.}
  \label{fig:gem5_eval}
\end{figure}

\sloppy
Figure \ref{fig:gem5_eval} shows the average throughput of TuRaN for SPEC2006 workloads~\cite{henning2006spec}. We calculate the throughput by finding the time it takes to generate random numbers in caches' idle cycles. 

TuRaN generates true random numbers in the L1 data cache (L2 cache) with an average throughput of $4.03Gbps$ ($10.95Gbps$) and a maximum throughput of $9.96Gbps$ ($13.46Gbps$). Since TuRaN evicts the previously-identified cache line whenever it generates a true random number, TuRaN degrades the system performance with an average of $4.86\%$ ($1.92\%$).  Since we use the CPU to perform the SHA-256 function with the throughput of $27.984Gbps$, the hash function does not reduce the throughput of TuRaN.

Our system integration has a negligible area overhead. We evaluate the area overhead of TuRaN using CACTI~\cite{muralimanohar2009cacti6}. The overhead of Drowsy Cache implementation on L1 and L2 Cache, which is lower than 3\% for each cache line reported on prior work~\cite{flautner2002drowsy}, is $0.00135mm^2$, $0.0108mm^2$ respectively. For the 1024-bit buffer, $r_{random}$, it is $0.0003 mm^2$. TuRaN requires an additional $0.00165mm^2$ for L1 data cache integration and $0.0111mm^2$ for L2 cache integration. 

\subsection{Discussion}
\label{subsec:sysint-dis}
In this study, we focus on how to leverage widely available SRAM devices to become a promising TRNG that can be used in computing devices of all scale. In this section, we discuss 1) how practical it is to implement TuRaN into commodity systems, 2) how to minimize TuRaN's performance overhead, 3) hardware-software interface of TuRaN, 4) TuRaN's the robustness and reliability, and 5) the integration of TuRaN into resource-constrained devices.

\noindent
\textbf{Practicality of TuRaN:} Recent Intel CPUs (e.g., VccCache in Intel CPUs~\cite{hammarlund2014haswell}) provide support for voltage underscaling in its caches. TuRaN can be implemented in these processors without any hardware modifications. Such an implementation requires evicting all cache blocks in the cache where TuRaN is implemented (e.g., L1 cache) before the true random number generation process starts. This is because existing CPUs provide coarse-granularity voltage control in caches, i.e., scaling the voltage of the cache affects all cache blocks in the cache. Evicting all cache blocks in the L1 cache would induce considerable system performance overhead as the L1 cache is not available to the processor until the random number generation process finishes. To enable the state-of-the-art SRAM-based TRNG~\cite{zhang2020improved} mechanism in the off-the-shelf processors, it needs 1) voltage manipulation and 2) physical preprocessing steps (e.g. irradiation exposure) which make the state-of-the-art SRAM-based TRNG depend on modifications on top of off-the-shelf processors. TuRaN would be more viable than the SRAM-based TRNGs as it only requires voltage manipulation. Regardless, TuRaN still outperforms aforementioned SRAM-based TRNGs 7.48x in throughput, 5.39x in latency, and 5.09x energy consumption.

\noindent
\textbf{Predicting Idleness in Caches:} To reduce the interference with the concurrently running applications and minimize the performance overhead, TuRaN leverages the idle time periods in caches to generate random numbers. Even though the length of an idle interval is not known in advance, it can be predicted using memory addresses that are being accessed and the occupancy in load-store queues. Recent work~\cite{bostanci2022dr}, developed an end-to-end system design for DRAM-based TRNGs that predicts the length of idle intervals in DRAM using last accessed memory addresses and the number of requests in memory request queues. We can use a similar approach for our architecture-level design and predict idle intervals in the selected cache level that are long enough to generate random numbers without any overhead by monitoring the occupancy in load-store queues and the last hit/miss memory addresses. If there are no idle cycles in the cache, the cache controller can either stall the memory requests and generate random numbers until the random number buffer is full or use a more sophisticated policy to minimize the unfairness induced by true random number generation and performance degradation of concurrently running applications. However, even with the workloads that utilize the cache bandwidth the most, we observe that there are sufficiently long idle intervals available for random number generation. As an additional countermeasure to this, all cache levels can be used to generate random numbers to create more opportunities for random number generation.  

\noindent
\textbf{The HW/SW Interface:} Various HW/SW interfaces can be used to enable TuRaN on modern systems, including but not limited to I/O buses and ISA extensions. Using memory-mapped space I/O datapaths to provide a simple interface to read the entropy buffer, $r_{random}$ which already in-used in modern systems to retrieve random number to the processor (i.e., TRNG\_OUT in AMD~\cite{amdrng} or APB-based slave interface in ARM~\cite{armtrng}). Another approach can be adding a new ISA instruction to read the $r_{random}$ and send the random number to the processor which is also employed in the modern systems (e.g. Intel RDRAND instruction\cite{inteldrng}).

\noindent
\textbf{Robustness and Reliability:} TuRaN is resilient against attacks that exploit process, voltage, and temperature variation that reduce TRNG entropy. To prevent this type of attacks, modern processors are already equipped with hardware that performs TRNG robustness and self-validation tests (e.g. Intel’s Online Health Tests and Built-in Self Tests~\cite{inteldrng}). These tests enable processors to track the entropy in the output of a TRNG. If the harvested entropy is not sufficient, a processor does not use the TRNG output. Thus, attackers cannot manipulate the output of a TRNG by controlling environmental parameters directly or indirectly.

\noindent
\textbf{Integration of TuRaN on Low-End Devices:} TuRaN can be integrated into resource-constrained microcontrollers that do not have dedicated TRNG hardware in two ways. First, entropy generated by performing voltage underscaling to an SRAM cache line can be post-processed using SHA-256. SHA-256 would be performed by the microcontroller. Even the low-end microcontrollers that do not have the budget to implement dedicated hardware TRNGs (e.g., ARM-Cortex M0) already support SHA-256 operation with throughputs exceeding 1.6Kbps~\cite{sharkssl}. Thus, using SHA-256, TuRaN can provide substantial TRNG throughput in resource-constrained devices that cannot afford dedicated TRNG hardware. Second, true random numbers can be directly retrieved from SRAM cells without any post-processing (described in Section~\ref{subsec:quality}). A low-performance microcontroller can directly read these SRAM cells to generate true random bitstreams at higher throughput (compared to performing post-processing), avoiding the performance and energy costs of performing relatively complex SHA-256 operations.
\section{Related Work}\label{sec:related-work}
To the best of our knowledge, this is the first study to exploit undervolting-based timing faults on SRAMs to generate true random numbers. In Section \ref{subsec:comparison}, we extensively describe and compare two state-of-the-art SRAM-based TRNGs to TuRaN. In this section, we briefly describe other prior SRAM-based TRNGs and other memory-based (non-SRAM-based) TRNGs.
\subsection{SRAM-based TRNGs}
SRAM-based TRNGs are firstly proposed in \cite{holcomb2008power} by exploiting SRAM start-up values to generate true random numbers. However, a very little portion of SRAM behaves randomly and does not exceed 0.1 minimum entropy. Therefore, many other prior works~\cite{vanderLeest2012,zhang2020improved,kiamehr2017leveraging,rahman2016enhancing,sadhu2020sstrng,wang2020aging,yeh2019self,holcomb2007initial,clark2018sram,wang2020long,li2015pufkey} propose SRAM-based TRNG to achieve higher minimum entropy and high-proportion of randomness in SRAMs.~\cite{vanderLeest2012} proposes an efficient algorithm to generate SRAM-based RNG by using two-stage post-processing functions, SHA-256 and deterministic random bit generator (DBRG). The resulting bitstream is used as a seed to generate pseudo-random numbers. To increase minimum entropy, prior work~\cite{kiamehr2017leveraging} leverages transistor aging impact. Similar to transistor aging, \cite{rahman2016enhancing} proposes a noise-sensitive embedded SRAM(NS-SRAM)-based TRNG to increase minimum entropy and quality of random numbers. However,~\cite{rahman2016enhancing} does not take into account their post-processing function, to evaluate throughput and area. 

Every prior SRAM-based TRNG, \textit{(i)} cannot maintain continuous operation, \textit{(ii)} has low-throughput at high latency due to their power-up cycle dependence, \textit{(iii)} can not achieve energy efficiency since they operate at nominal operating parameters, and \textit{(iv)} can not easily be implementable on commodity devices due to the \textit{(i)} and \textit{(ii)}.
\subsection{Non-SRAM-based TRNGs}
\noindent
\textbf{DRAM.} Prior works on DRAM-based TRNG use different approaches to generate true random numbers, such as intentionally violating the DRAM timing parameters\cite{olgun2021quac,talukder2019exploiting,kim2019d,sutar2016d,keller2014dynamic} and using start-up values\cite{eckert2017drng,tehranipoor2016robust}. These proposals either (i) do not consider energy consumption or (ii) are not energy efficient, such as TuRaN $37.6x$ times consume lower energy than the best energy efficient DRAM-based TRNG\cite{kim2019d} (iii) or can not achieve high-throughput.  

\noindent
\textbf{FLASH.} Prior Flash-based TRNG proposals\cite{ray2018true,chakraborty2020true,wang2012flash} exploit the thermal noise and RTN to generate true random numbers. However, the highest throughput among these proposals is 1Mbps, $1812x$ times lower than TuRaN.

\noindent
\textbf{Existing TRNGs in Commodity Systems.}
Commodity off-the-shelf systems use dedicated TRNG hardware~\cite{inteldrng,amdrng,armtrng} to ensure security-critical operations. These TRNGs in commodity processors have limited throughput. The entropy source of Intel Ivy Bridge's TRNG~\cite{inteltrng}, achieves 3Gbps throughput which is 1.34x and 3.32x smaller than TuRaN's L1D cache integration throughput and L2 cache integration throughput, respectively. Dedicated TRNG hardware used in ARM~\cite{armtrng} can produce 10Kbps of entropy when the core runs at 200MHz which is 18912x lower than TuRaN's entropy.
\section{Conclusion}
In this study, we introduce TuRaN, an energy-efficient SRAM-based TRNG with high-throughput at low latency that can be implemented in modern systems at low cost. TuRaN exploits supply voltage underscaling on SRAMs and post-processes the resulting timing faults with the SHA-256 hash function to generate true random numbers. We characterize and evaluate TuRaN on two identical FPGA boards. We show how frequency, voltage level, and data pattern affect entropy. We evaluate the random numbers generated by TuRaN in terms of quality, throughput, energy, and latency. We show that TuRaN generates random numbers that pass \emph{all} the NIST STS test with the throughput of $1.6Gbps$ on average, energy of $0.11nJ$ per true random bit, and the latency of $278.46\mu s$. TuRaN significantly outperforms the state-of-the-art SRAM-based TRNGs in throughput by $2.26x$, energy efficiency by $5.09x$, and latency by $5.39x$. We demonstrate two potential integration of TuRaN in a state-of-the-art CPU L1 data cache (and L2 caches) and, we achieve $4.03Gbps$ ($10.95Gbps$) throughput on average with a negligible overhead of $0.00165mm^2$ ($0.0111mm^2$).


\bibliographystyle{IEEEtranS}
\bibliography{refs}

\end{document}